\documentclass[aip,cha,onecolumn,reprint,
]{revtex4-2}
\usepackage[utf8]{inputenc}
\usepackage{bm}
\usepackage{amssymb}
\usepackage{mathtools}
\usepackage{amsmath}
\usepackage{xcolor}
\usepackage{enumitem}
\usepackage{natbib}
\usepackage{hyperref}
\usepackage{graphicx}
\usepackage{ragged2e} % for \justifying
\usepackage[normalem]{ulem}
\usepackage{gensymb}

\usepackage{cellspace} 
 \hypersetup{colorlinks=true, linkcolor=black, breaklinks=true, urlcolor=black, citecolor=black}
\usepackage{lipsum}
\usepackage[ruled,vlined]{algorithm2e}

\def \bu {\pmb{u}}

\def \bx {\pmb{x}}
\def \by {\pmb{y}}

\def \bz {\pmb{z}}

\def \be {\pmb{e}}

\def \Tr {\text{Tr}}
\usepackage{rotating}
\usepackage[english]{babel}
\usepackage{amsthm}

\theoremstyle{definition}
\newtheorem{definition}{Definition}

\newtheorem{example}{Example}

\usepackage{subfig}

\makeatletter
\renewcommand{\fnum@figure}{FIG. \thefigure}
\makeatother

\makeatletter
\providecommand{\@LN}[2]{}
\makeatother

\makeatletter
\long\def\@makecaption#1#2{%
  \vskip\abovecaptionskip
  \sbox\@tempboxa{#1: #2}%
  \ifdim \wd\@tempboxa >\hsize
    {\justifying #1: #2\par}
  \else
    \global\@minipagefalse
    \hb@xt@\hsize{\justifying\hfil #1: #2\hfil}%
  \fi
  \vskip\belowcaptionskip}
\makeatother

\usepackage{comment}
\usepackage[normalem]{ulem}  % provides \sout

\newcommand{\redstrike}[1]{%
  \bgroup
  \markoverwith{{\rule[0.6ex]{2pt}{0.8pt}}}%
  \ULon{#1}%
  \egroup
}

\begin{document}
%\title{Slow component resilience in timescale separated synchronized oscillators} 

\title{ 
The Frequency Response of  Networks as Open Systems
% \\ Modulates  Environmental Signal Transmission
}
%\title{Passing vs blocking behavior in the frequency response of networked systems}
\author{Amirhossein Nazerian}
\affiliation{Department of Mechanical Engineering, University of New Mexico, Albuquerque, NM 87131, United States of America}
\author{Malbor Asllani}
\affiliation{Department of Mathematics, Florida State University, 1017 Academic Way, Tallahassee, FL 32306, United States of America}
\author{Melvyn Tyloo}
\affiliation{Living Systems Institute, University of Exeter, Exeter, EX4 4QD, United Kingdom} 
\affiliation{Department of Mathematics and Statistics, Faculty of Environment, Science, and Economy, University of Exeter, Exeter, EX4 4QD, United Kingdom}
\author{Wai Lim Ku}
\affiliation{Center for Applied Data Science and Analytics, Howard University, Washington, DC 20059, United States of America 
}
\affiliation{Center for Sickle Cell Disease, Howard University, Washington, DC 20059, United States of America 
}
\affiliation{Department of Medicine, Howard University, Washington, DC 20059, United States of America 
}
\author{Francesco Sorrentino}
\email{fsorrent@unm.edu}
\affiliation{Department of Mechanical Engineering, University of New Mexico, Albuquerque, NM 87131, United States of America}
\affiliation{Max Planck Institute for the Physics of Complex Systems, 01187 Dresden, Germany}

% \begin{abstract}
%     Many biological, technological, and social systems are effectively networks of interacting individual systems. Typically, these networks are not isolated objects but interact with their environment through both signals and information that is received by specific nodes with input function or released to the environment by other nodes with output function. An important class of such signals that we study is sinusoidal and periodic, although our analysis encompasses arbitrary signals, including noise. A fundamental question is whether the structure of different networks, together with the particular selection of input and output nodes, is such that it favors the passing or blocking of such signals. For a given network and a given choice of the input and output nodes, a natural and general quantification of the extent to which input signals are amplified by the network is given by the H2-Norm. We analyze several empirical datasets and conjecture that real networks have evolved to achieve a desired passing or blocking behavior, as evidenced by several empirical networks in this study. 
%     \textcolor{teal}{
%     Based on our study, Food webs, Genetic networks, and Pathway networks show passing behavior, while Power Grids often show blocking behavior. 
%     }
% \end{abstract}

\begin{abstract}
\textcolor{black}{
Many biological, technological, and social systems can be effectively described as networks of interacting subsystems. Typically, these networks are not isolated objects, but interact with their environment through both signals and information that is received by specific nodes with an input function or released to the environment by other nodes with an output function. 
An important question is whether the structure of different networks, together with the particular selection of input and output nodes, is such that it favors the passing or blocking of such signals. 
For a given network and a given choice of the input and output nodes, the $\mathcal{H}_2$-norm provides a natural and general quantification of the extent to which input signals—whether deterministic or stochastic, periodic or arbitrary—are amplified.
{We analyze a diverse set of empirical networks and find that many naturally occurring systems—such as food webs, signaling pathways, and gene regulatory circuits—are structurally organized to enhance the passing of signals;
in contrast,  the structure of engineered systems like power grids appears to be intentionally designed to suppress signal propagation.}
%facilitating the efficient flow of biomass, information, or regulatory activity. %This passing behavior culminates in directed acyclic graphs (DAGs), for which we analytically show that amplification depends on the number and length of input-output pathways, which is consistent with the well-known tendency of naturally emerging networks to approximate DAG structures.
%In contrast, the structure of engineered systems like power grids appears to be intentionally designed to suppress signal propagation.%, reflecting the need for precise regulation of transmitted signals, such as voltage phase differences, to maintain synchronized operation.
}
\end{abstract}

%%%%%%%%%%%% Shortened version abstract %%%%%%%%%%%%%%
% Biological, technological, and social systems can be modeled as networks of interacting subsystems that exchange signals with their environment through designated input and output nodes. A central question is whether a network’s structure, together with the chosen input and output nodes, promotes signal transmission or blocks it. For a fixed network and input output selection, the $\mathcal{H}_2$-norm quantifies how input signals, whether deterministic or stochastic and whether periodic or arbitrary, are amplified. Examining diverse empirical networks, we conjecture that many natural systems such as food webs, signaling pathways, and gene regulatory circuits are structurally organized to enhance signal passing and enable efficient flow of biomass, information, or regulatory activity. This passing behavior culminates in directed acyclic graphs (DAGs), for which we show analytically that amplification depends on the number and lengths of input output pathways. By contrast, engineered networks such as power grids appear designed to suppress propagation to support synchronized operation.

\maketitle

\section{Introduction}

Networked dynamical systems, composed of  are omnipresent is nature and engineering, ranging from neuronal dynamics to electric power grids. They consist of many interacting units, each with its own intrinsic parameters and degrees of freedom, whose mutual coupling typically gives rise to rich collective behaviors. 
Usually, these networks are not isolated from their environment, but are subjected to external perturbations, operational changes, and external stimuli. 
Examples are neuron networks subject to sensory stimuli \cite{horrocks2024flexible,xue2024theoretical}, power grids that both absorb power produced at the generators and deliver power at the loads \cite{sajadi2022synchronization,martinez2025transmission,fritzsch2024stabilizing}, and networks subject to the effects of environmental noise, such as ecological networks \cite{meng2020tipping}. At the same time, networks produce outputs that affect the environment. Hence, the response of networked systems to external inputs essentially depends on the units chosen as input nodes and output nodes and the structure and characteristics of the network connectivity.

{While a large literature has studied control strategies for networks, see e.g., \cite{yan2015spectrum,gao2014target,klickstein2017energy,klickstein2018control}, this paper focuses on the response of networked systems to various environmental inputs, not limited to control signals.}
% Reference \cite{tyloo2018robustness} addressed the global response of a network to external stimuli.
% In this study, the network was taken as symmetric, but similar results have been found for asymmetric and normal matrices \cite{young2010robustness}.
% The network response to single-node input signals was studied in \cite{tyloo2019key}. The connectivity matrix was asymmetric, but not the edge weights: the asymmetry resulted from the reduction from a second-order ODE system to a first-order ODE system.
% Reference \cite{tyloo2023finite} investigated how noise properties are transmitted from a single and multiple input nodes to all the others in the network. The interactions are symmetric, but the connectivity matrix is asymmetric due to the second to the first order ODE transformation.
% Reference \cite{zhang2019fluctuation} investigated the frequency response of power grid models, i.e., asymmetric connectivity matrix, because of the second-order ODE, but still, the interactions are symmetric.
% Reference \cite{lizier2023analytic} evaluated whether a directed network is synchronizable and assessed the quality of the synchronized state by calculating the amplitude of the excursion following a perturbation. They expressed the steady state variance as the sum over the products of weighted path between every nodes in the network and also observed that shorter paths contribute the most to the variance. 
References \cite{tyloo2018robustness,young2010robustness,tyloo2019key,tyloo2023finite,zhang2019fluctuation,lizier2023analytic} have studied the response of a networked system to external stimuli. 
% These works focused on graphs with symmetric adjacency matrix or, in the most general case, asymmetric but normal adjacency matrices.
% In our work, we do not assume the connectivity to be symmetric, and we also focus on the important class of sinusoidal and periodic inputs, although our treatment encompasses arbitrary signals and noise.
%\color{black}
{These studies have primarily focused on graphs with symmetric connectivity or, at most, on asymmetric but normal adjacency matrices. Yet, most real-world networks are highly non-normal, strongly directed, and often exhibit pronounced hierarchical organization---structural features that profoundly shape their dynamical response \cite{johnson2017looplessness, asllani2018structure, asllani2018topological, MUOLO2019Patterns, muolo2020synchronization, o2021hierarchical, duan2022network, nazerian2023Communications, ramon2024entropy}.
{ In such networks, non-normality acts as an algebraic expression of directedness and emerges from strong asymmetries in the adjacency pattern with a clear source and sink structure, and induces strongly biased directions of flow. Structurally, highly non-normal networks are close to acyclic, with a dominant directed acyclic backbone and only sparse or weak cycles; this can be quantified, for example, by measuring the distance from the largest directed acyclic subgraph, a quantity that indicates how close the network is to a loopless structure~\cite{johnson2017looplessness, asllani2018structure, asllani2018topological}.} 

{
Consistent with this view, recent work shows that increasing directedness and non-normality leads to flows that are more localized, strongly biased downstream, and effectively irreversible, so these architectures naturally favor one way signal transfer toward terminal components and sink nodes~\cite{o2021hierarchical, ramon2024entropy}.}
Motivated by these properties, our approach does not assume symmetry and instead addresses the response to general input signals, such as sinusoidal, periodic, and stochastic, within a unified framework.}
{Recent works have focused on how different nonlinearities affect the propagation patterns and stability in complex networks~\cite{hens2019spatiotemporal,meena2023emergent}. Here, instead, we aim at characterizing whether specific types of networks tend to facilitate or block signal transmission, in terms of the combined effect of network structure as well as the corresponding dynamics.
}

%Reference \cite{tyloo2018robustness} addressed the global response of a network to external stimuli. In this study, the network was taken as symmetric, but similar results have been found for asymmetric and normal matrices \cite{young2010robustness}. The network response to single-node input signals was studied in \cite{tyloo2019key}. The connectivity matrix was asymmetric, but not the edge weights: the asymmetry resulted from the reduction from a second-order ODE system to a first-order ODE system. Reference \cite{tyloo2023finite} investigated how noise properties are transmitted from a single and multiple input nodes to all the others in the network. The interactions are symmetric, but the connectivity matrix is asymmetric due to the second to the first order ODE transformation.
%Reference \cite{zhang2019fluctuation} investigated the frequency response of power grid models, i.e., asymmetric connectivity matrix, because of the second-order ODE, but still, the interactions are symmetric.
%Reference \cite{lizier2023analytic} evaluated whether a directed network is synchronizable and assessed the quality of the synchronized state by calculating the amplitude of the excursion following a perturbation. They expressed the steady state variance as the sum over the products of weighted path between every nodes in the network and also observed that shorter paths contribute the most to the variance.

\begin{figure*}
    \centering
    \includegraphics[width=0.7\linewidth]{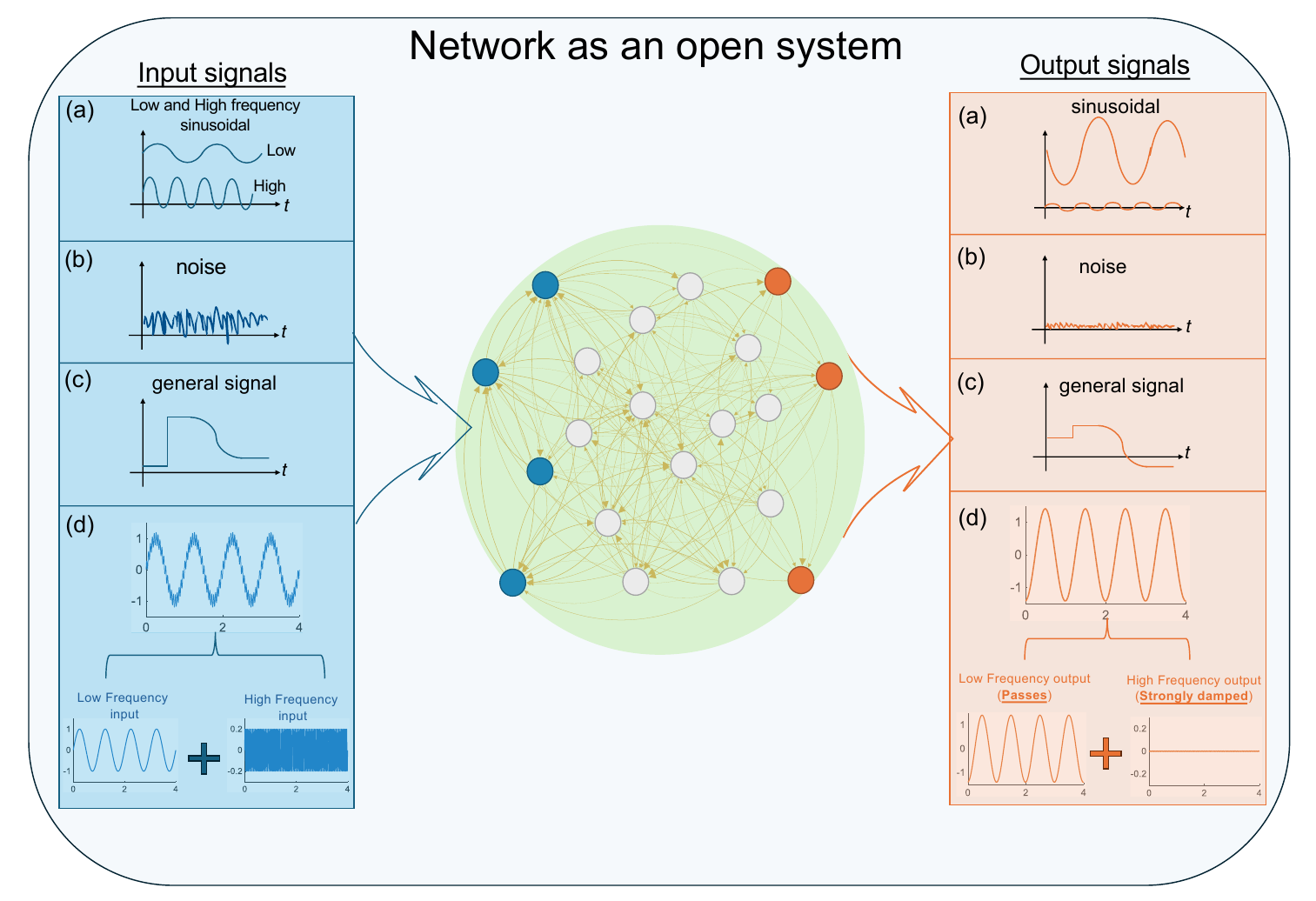}
    \caption{ {
    \textbf{Networks as open systems.} The schematic summarizes the goal of the paper in studying how an external signal is altered when it passes through a network. 
    The network is subjected to various input signals from the environment, which are only received at the input nodes (in black.)
    The response of the network is measured at some output nodes (in orange). Several input signals that are considered in this paper and corresponding output signals are  shown. In (a)
    a low-frequency sinusoidal input signal may experience a phase shift and amplification, while a high-frequency one may be damped. 
    The case in (b) is that of a network that suppresses noise. In (c) the shape of a general output signal may not be the same as its corresponding input signal.
    In (d) the input signal contains both a low frequency and a high frequency components. In the output signal, the low frequency passes while the high frequency is strongly damped and thus effectively blocked.}}
    \label{fig1}
\end{figure*}

\color{black}
This paper examines how complex networks of coupled dynamical systems respond to environmental interactions, where external signals are received at designated input nodes and measured at specific output nodes~\cite{liu2011controllability,yuan2013exact,wang2017physical,liu2016control}. An open network is defined by (i) its internal topology i.e., the set $\mathcal N$ of the network nodes and the set $\mathcal E$ of directed, possibly weighted, edges that connect them, (ii) the set of input nodes $\mathcal I \subseteq \mathcal N$ where external stimuli are applied, and (iii) the set of output nodes $\mathcal O \subseteq \mathcal N$ where the response is measured. As illustrated schematically in  Fig.~\ref{fig1}, the network structure and its internal connectivity shape how input signals—including sinusoidal, periodic, and noisy—are transformed as they propagate from input to output, with different frequency components affected by amplification, attenuation, or distortion, depending on the structural properties of the network.  Such signals are ubiquitous in natural and technological systems—for example, gene regulatory networks responding to circadian rhythms~\cite{yan2008analysis}, power grids operating at standardized frequencies of 50--60\,Hz~\cite{machowski2020power}, and neuronal circuits processing signals across a broad frequency range of roughly 0.1--100\,Hz~\cite{terlouw2016consciousness}—and we emphasize that our framework also applies to noise and arbitrary signals.

{
In panel (a) of Fig.~\ref{fig1}, a low-frequency sinusoidal input may undergo phase shift and amplification, whereas a high-frequency input is attenuated. Panel (b) illustrates a network that suppresses noise. In panel (c), the output signal generally differs in shape from its corresponding input. In panel (d), the input contains both low- and high-frequency components; in the output, the low-frequency component passes through, while the high-frequency component is strongly attenuated and effectively filtered out. The Methods Section \ref{methods:example} presents a numerical example of signal transformation when it passes through a real connectome network.} % The input signal is composed of a low-frequency signal and a high-frequency signal.The input signal enters the network through the black node and the output signal is measured at the red node. The output signal has the same two frequencies as the input signal, but with different amplitudes and phases.}

{ In this paper, we adopt a frequency-response viewpoint: we represent each system through its transfer function and quantify signal amplification using the $\mathcal{H}_{2}$-norm, which captures responses to sinusoidal, periodic, noisy, and more general inputs.}
 Our analysis investigates how the topology of a network and the placement of its input and output nodes shape the amplification or attenuation of external signals. 
 %We characterize this behavior through the network’s transfer function, or frequency response, and use the $\mathcal{H}_{2}$-norm as a universal metric that quantifies signal amplification---capturing responses to sinusoidal, periodic, noisy, and general inputs. 
 We establish a direct relationship between the network’s topology, its transfer function, and the controllability Gramian, which serves as a central tool in our study. This framework allows us to assess whether empirical networks tend to amplify (\emph{pass}) or suppress (\emph{block}) external inputs for given input-output configurations.

Building on this framework, { we next apply it to a broad collection of empirical networks from biology, technology, and social systems in order to uncover structural principles governing how they transform signals from input to output.} %we perform a broad analysis of empirical networks from different fields, to uncover structural principles governing their signal transformation properties.
Key findings are that many biological networks—such as food webs, signaling pathways, and gene regulatory networks—tend to exhibit architectures that enhance the passing of signals, supporting the efficient flow of biomass, information, or regulatory activity. This signal-passing behavior is especially evident in directed acyclic graphs (DAGs), { namely directed networks that contain no directed cycles,} which are known to closely approximate many natural real-world networks \cite{asllani2018structure} and { suggests} that such architectures inherently promote signal propagation. Motivated by this, we analytically compute the DC gain {i.e., the input to output gain when the input is a constant signal} for DAG networks, { thereby linking amplification or attenuation directly to the number and length of the input–output pathways.} %demonstrating that the amplification or attenuation is determined explicitly by the number and length of the input-to-output pathways. 
In contrast, engineered systems such as power grids—often modeled as mostly undirected (symmetric) graphs—appear to be purposefully designed to suppress signal amplification, reflecting the need for regulation of transmitted signals, such as voltage phase differences, to ensure stable and synchronized operation.

% % Figure \ref{fig1} demonstrates the general theme of the paper in studying the network response to external stimuli and the transformation of the input signals in open networks.
% % The illustration is intended to highlight that input signals of different frequencies may experience different transformations as they pass through the network. 
% % We will also study the response of systems to noise and general signals.
% We derive relationships between the network topology, the network transfer function, and its controllability Gramian. 
% {Key findings include the role of shortest unweighted paths in determining the frequency response, as well as the important connection between non-normality of the network topology and passing vs blocking behavior. 
% Empirical studies on real-world networks (e.g., power grids, food webs, pathway networks, and gene regulatory networks) reveal that some networks are optimized for passing signals, while others favor blocking.  
% %Overall, the work bridges theoretical insights with practical applications, offering tools to design or modify networks for desired signal processing behaviors.
% }

\section{Results}

\subsection{Frequency Analysis of Networks as Open Systems}

Many natural and engineered systems—such as gene networks, power grids, and neural circuits—can be modeled as open networks, where each node evolves through intrinsic dynamics, interacts with neighbors, and communicates with the environment via inputs and outputs. While inherently nonlinear, such systems can often be approximated near equilibrium by a linear time-invariant (LTI) model, enabling powerful analysis through linear control theory.

% {Malbor and all, I propose that we simplify things here a bit. In the paper `Network structural origin of instabilities in large complex systems' all the authors say is: Given that many real network systems operate near an equilibrium, we consider the class of (nonlinear) systems whose linearization around a given reference equilibrium state is described by \( \dot{x}_i = - \alpha x_i + \sum_j A_{ij} x_j \). I think we need something like that.} We consider scalar nodal dynamics of the form \( \dot{x}_i = f_i(x_i) + \sum_j A_{ij} g_{ij}(x_i, x_j) \) in a networked system, where each node \( i \) evolves under intrinsic dynamics \( f_i \), and interacts with neighbors \( j \) through a weighted coupling function \( g_{ij} \) defined by the adjacency matrix \( A \). Assuming linear coupling and linearizing around a stable fixed point \( x_i^* \) with differentiable, monotonic \( f_i \) yields the Jacobian \( A + \mathrm{diag}(f_i'(x_i^*)) \).

{Throughout this work we consider a general nonlinear model for the network dynamics of the form,
$   \dot{z}_i(t) = f(z_i(t)) + \sigma \sum_j \tilde{A}_{ij}\, g(z_i(t),z_j(t))$,
$i=1,..,N$, where $z_i(t) \in \mathbb{R}$ is the state of system $i$,
the function \(f : \mathbb R \rightarrow \mathbb{R}\) describes the intrinsic dynamics of each node, \(g : \mathbb R \times  \mathbb R \rightarrow \mathbb{R}\) the pairwise interactions, the scalar \(\sigma\) is the global coupling strength, and \(\tilde{A} \in \mathbb{R}^{N \times N} \) is the directed adjacency matrix of the network. 
With suitable choices of the functions \(f\) and \(g\), this general framework has been applied to model diverse networked systems, such as power grids, food webs with competitive and mutualistic interactions, gene-regulatory networks, and neuronal population dynamics.
Since many real-world networked systems operate near a stable equilibrium, we study here the linearized dynamics around such an equilibrium,
$\dot{x}_i = -A_{ii} x_i + \sum_j A_{ij} x_j, i = 1,\ldots,N,$ 
where the interdependencies among the linearized states \(x_i\) of the network nodes are captured by the network Jacobian matrix \(A = [A_{ij}]\). Since here we consider a stable fixed point, all eigenvalues of the Jacobian matrix $A$ have negative real part.
For more details on the steps involved in the linearization, see the Methods Section \ref{methods:A}. We remark that the network Jacobian $A$ shares a similar structure as the network adjacency matrix $\tilde{A}$, which is demonstrated with an example in the Methods Section \ref{methods:A}.

%{\sout{ Throughout this work, we  ensure asymptotic stability by  applying the spectral shift $A \leftarrow A - c \mathbb{I}$,  which moves all eigenvalues of $A$ into the left half of the complex plane, while leaving the eigenvectors unchanged.}

\color{black}
This formulation leads to a networked system governed by the standard LTI equations:
\begin{equation} \label{eq:continuous}
\begin{aligned}
    \dot{\bx}(t) &= A \bx(t) + B \bu(t), \\
    \by(t) &= C \bx(t),
\end{aligned}
\end{equation}
where \( \bx(t) \in \mathbb{R}^N \) is the network state vector, and \( \bu(t) \in \mathbb{R}^m \) is the vector of external input signals acting on the network, reflecting its open nature. The output vector \( \by(t) \in \mathbb{R}^p \) collects signals from selected nodes. 
{The matrix $A$ is the network Jacobian introduced before, %includes information on the node-to-node interactions (off diagonal entries) and on the  individual dynamics at each node (diagonal entries.) 
the matrices $B$ and $C$ encode information on the nodes where inputs enter the network and the nodes from which outputs are measured, respectively. Each column of $B$ corresponds to one input signal and contains a single 1. If $B_{ij} = 1$, this means that \emph{input $j$ is received at node $i$}.
Similarly, each row of $C$ corresponds to one output signal and also contains a single 1. If $C_{ij} = 1$, this means that \emph{output $i$ is measured at node $j$}.
In short, each input is injected at exactly one node, and each output is read from exactly one node.}

% The matrices $B$ and $C$ are assumed
% to have versors as their columns and rows, respectively.
% That is, if $B_{ij} = 1$, then $B_{kj} = 0$, $\forall k \neq i$.
% Similarly for the matrix $C$, if $C_{ij} = 1$, then $C_{ik} = 0$,
% $\forall k \neq i$. Our choice of the matrices $B$ and $C$ describes
% which nodes are input and output nodes, respectively. If
% $B_{ij} = 1$, it means that node $i$ is an input node, $i \in \mathcal I$ and receives
% input signal $j$. Similarly, if $C_{ij} = 1$, it means that
% node $j$ is an output node, $j \in \mathcal O$ and transmits output signal $j$.

% The matrices \( B \) and \( C \) define which nodes are controlled and observed, respectively, and are taken to consist of canonical basis vectors, to ensure that input and output act on individual nodes.
% \textcolor{teal}{
% That means, the columns of the matrix $B$ (the rows of matrix $C$) are basis vectors $\be_i$ where the only non-zero entry in $\be_i$ is entry $i$ and is equal to $1$ (where the only non-zero entry in $\be_j^\top$ is entry $j$ and is equal to $1$).
% }

The LTI approximation provides a principled and analytically tractable framework for studying networked systems operating near stable fixed points, during short transients, or under weak perturbations, where linearized dynamics capture the dominant system behavior. 
This includes many biological and engineered systems functioning under steady-state conditions. 
Also systems near criticality, such as second-order phase transitions, where linear responses captures essential behavior \cite{binney1992theory}, can be modeled as LTI systems. 
Even in synchronization problems, governing models in neurodynamics, the LTI framework remains valid when the system is close to the onset of collective oscillations, which typically arise through a Hopf bifurcation occurring at a specific mode (e.g., zero wavenumber or the zero coupling eigenvalue) \cite{kuramoto1984chemical, cross2009pattern, asllani2025pattern}. 
In such cases, the dynamics can be linearized around the fixed point prior to bifurcation, yielding time-invariant Jacobians that characterize local behavior. 
%{This formulation aligns conceptually with the structural controllability framework introduced by Liu et al.\cite{liu2011controllability},\footnote{To avoid overstating its generality, our approach is narrower and more principled—focusing specifically on systems linearized near equilibrium and leveraging classical transfer function analysis.}. }
Crucially, we find an important relation with the non-normality of the network {Jacobian} matrix, which is known to induce transient amplification and shape control energy\cite{higham2005spectra} and is an ubiquitous feature of real-world networks. %This is especially true for systems with directional flow and hierarchical organization~\cite{asllani2018structure,o2021hierarchical}, supporting the expectation that many naturally occurring networks exhibit passing behavior when inputs are suitably placed.
{Finally, one should note that many-body interactions can also be accounted for within our framework. Indeed, when linearizing the dynamics around a stable fixed point, the interaction can, at leading order, be described by an effective pair-wise directed network through the Jacobian. In what follows, we restrict our attention to this first-order (linear) approximation, although higher-order terms e.g. in phase-reduction expansions can generate explicit many-body effective interactions beyond this description~\cite{leon2019phase}.}

In this work, we adopt a frequency-domain approach to characterize the dynamics of networked open systems. Any periodic input can be decomposed into a superposition of sinusoids via Fourier analysis, enabling the study of the system’s frequency response to generic inputs. For the LTI model in Eq.~\eqref{eq:continuous}, the transfer function {\cite{ogata2010modern}} is defined as
\begin{equation} \label{eq:transferfunction}
    G(s) = C(s\mathbb{I} - A)^{-1}B. 
\end{equation}
By evaluating {the transfer function} at \( s = \jmath \omega \), where \( \omega \) is the input frequency and \( \jmath = \sqrt{-1} \), we probe the system’s steady-state response to sinusoidal inputs. {After the transient has elapsed, the} output oscillates at the same frequency \( \omega \) {of the input signal}, but with altered magnitude and phase, given by
\begin{equation} \label{Eq:mag_phase}
    \text{Magnitude} = 20 \log_{10} |G(\jmath\omega)|, \qquad \text{Phase} = \angle G(\jmath\omega).
\end{equation}
{
Specifically, if the input signal $u(t) = \sin (\omega t)$, then the output signal $y(t) = \text{Magnitude} \cdot \sin(\omega t + \text{Phase})$
\cite{ogata2010modern}.} {The factor \(20\log_{10}(\cdot)\) in Eq.\ \eqref{Eq:mag_phase} expresses the magnitude \(|G(j\omega)|\) in decibels (dB), the conventional logarithmic unit for system gain.
}
This frequency-based formulation underpins our analysis and enables principled insights into how the network structure shapes signal transmission and network performance.

An important case of the frequency response analysis arises in the low-frequency limit, which corresponds to the case of slowly varying or constant inputs, such as step functions \cite{Shmaliy2007}. This steady-state behavior is captured by the so-called DC gain. In this limit, the frequency variable approaches zero, \( \omega \to 0 \). 
We have provided a general formulation for the DC gain in Methods \ref{sec:dcgain}.

\color{black}

%{Let's assume that we magically know the static gain. Can we obtain an approximation for the H2 norm by using the magnitude plot in Fig. 3B?}{Not necessarily. Static gain plays a role, but it is not dominant. As we discussed in the meeting on 6/11/2025, the placement of the zeros and the poles of the system greatly affects the value of the H2 norm. Even if we consider exact poles and assume there are no zeros in the system, the estimation still can be far off by orders of magnitude. } {Can we just obtain an approximation by computing the integral under the curve as two parts, a constant part equal to the gain up to the corner frequency, and a decaying part after that?  }{It cannot be done as you have it in mind. We can talk in a meeting, and I will show you the linear-linear plot vs log-log to see how it is difficult to integrate the linear-linear plot, although it looks simple in log-log.} {OK}

\subsection{$\mathcal{H}_2$-Norm}

% {The $\mathcal{H}_2$-Norm of a network as an open system, i.e., coupled to the environment, defined by the triplet $A,B,C$,  provides a universal measure of the input-to-output amplification provided by the network to input signals. As we will see, this is a broad metric with general relevance in at least three fundamental cases: that the input is an arbitrary time signal, that the input is a sinusoidal function with an arbitrary frequency, and that the input is a noisy signal.}

% The $\mathcal{H}_2$-norm of a system is a classical optimal control performance index that measures the input-output amplification \cite{Doyle1989State,Zhou1998Essentials}.
% In stochastically driven linear time-invariant systems, the $\mathcal{H}_2$-norm quantifies the variance amplification  \cite{JOVANOVIC20082090}.  
% {Variance amplification refers to how much the linear system amplifies the variance of random input disturbances, indicating the sensitivity of the system's output to stochastic fluctuations.}
% The $\mathcal{H}_2$-norm can also be considered as the energy of the response to the impulse input in deterministic systems \cite{Hassibi1999Indefinite}.
% In linear time-invariant systems, the $\mathcal{H}_2$-norm coincides with the trace of the output controllability Gramian. 

{ In this work, the $\mathcal{H}_2$-norm \cite{Zhou1998Essentials} serves as our principal analytical tool to quantify how the structure of a networked system shapes the amplification of inputs into outputs. For a given open network characterized by the triplet $A,B,C$, the $\mathcal{H}_2$-norm provides a universal and versatile measure of input-to-output amplification with general relevance in at least three fundamental scenarios: arbitrary time-varying inputs, sinusoidal inputs across the frequency spectrum, and noisy (stochastic) inputs. This classical optimal control performance index measures the total energy transfer from inputs to outputs~\cite{Doyle1989State,Zhou1998Essentials}, quantifies variance amplification in stochastically driven linear systems~\cite{JOVANOVIC20082090}, and coincides with the trace of the  Gramian in linear time-invariant settings~\cite{Doyle1989State,Zhou1998Essentials}. Intuitively, it reflects both the energy of the impulse response~\cite{Hassibi1999Indefinite} and the sensitivity of the system’s output to random input fluctuations.
}

The $\mathcal{H}_2$-norm squared for the system {with transfer function \eqref{eq:transferfunction}} measures the energy of the response to a unit impulse and can be computed as
\begin{align}
\begin{split}
    \| G \|_2^2 & := \dfrac{1}{2 \pi}\Tr\left[\int_{-\infty}^{+\infty} G(\jmath\omega)^H G(\jmath\omega) d \omega\right] \\
    &= \Tr \left[ \int_0^{+\infty} C e^{At} B B^\top e^{A^\top t} C^\top dt \right] = \Tr [ C W_c  C^\top ], \label{eq:H2norm}
\end{split}
\end{align}
where $W_c$ is the continuous-time controllability 
Gramian \cite{kailath1980linear} and is defined as
\begin{equation} \label{Eq:Gramian}
    W_c  = \int_0^{+ \infty}  e^{At} B B^\top e^{A^\top t} d t,
\end{equation}
{which, for simplicity, we refer to as the Gramian throughout the paper.}
{An important observation from Eq.\ \eqref{eq:H2norm} is that the $\mathcal{H}_2$-norm squared provides a measure of  amplification to all frequencies between $0$ and $\infty$, as well as to impulse time signals.} 

{The matrix $W_c$ in \eqref{Eq:Gramian} is symmetric and positive semi-definite and is finite if and only if the { Jacobian matrix $A$} has all eigenvalues with negative real parts. }
The matrix $W_c$ can be easily calculated by solving the following continuous-time Lyapunov equation \cite{kailath1980linear} for $W_c$:
\begin{equation}
    A W_c + W_c A^\top + B B^\top = 0.
\end{equation}
For more information on the properties of the Controllability Gramian \eqref{Eq:Gramian}, see Methods \ref{sec:propgramian}, 
which in particular discusses the important modularity property of the trace of the output  Gramian $\Tr(W_c^{out})$. Namely, for any choice of the sets of input nodes and output nodes, the 
trace $\Tr(W_c^{out})$ can be written as a sum of individual contributions, each one of which corresponds to a pair of input and output nodes (see Methods \ref{sec:propgramian}.) 
This has direct and important consequences, as it implies that if we assign the set of output nodes, a set of $k$ input nodes that minimizes (maximizes) $\Tr(W_c^{out})$ %, is simply to choose each node as an input node and record the resulting $\Tr(W_c^{out})$.With knowledge of the individual input node-output-node contributions, the choice of $k$ nodes that minimize/maximize $\Tr(W_c^{out})$ 
is simply composed of the $k$ nodes that provide the largest (smallest) individual contributions.
An analogous conclusion is obtained in the case that a given set of input nodes is assigned and one wants to select an optimal set of $k$ output nodes. 
{ Furthermore,} Supplementary Note~1 relates the output standard deviation to the $\mathcal{H}_2$-norm via the Gramian spectrum, and Supplementary Note~2 provides upper bounds for the {Bode} integral in terms of the eigenvalues and eigenvectors of the matrix $A$.

\begin{comment}
Specifically, we consider the system in Eq.\,\eqref{eq:continuous} with the transfer function
\begin{equation}
    G(s) = C (sI - A)^{-1}B.
\end{equation}
Frequency analysis studies how a sinusoidal input signal with unit amplitude transforms as it passes through the system. 
For this, we set the input to have frequency $\omega$, which corresponds to setting $s = \jmath\omega$, where $j = \sqrt{-1}$.
The output signal will have the same frequency as the input, $\omega$, while its magnitude and phase may be altered, i.e.,
\begin{equation}
    \text{Magnitude} = 20 \log_{10} | G(\jmath\omega) |, \quad \text{Phase} = \angle G(\jmath\omega).
\end{equation}
\end{comment}

{Up to this point, we have examined the overall network amplification of input signals across frequencies, as quantified  by the $\mathcal{H}_2$-norm. However, it is possible that a particular network will pass certain frequencies and block others, which can be studied by using Bode plots \cite{bode1945network}. { To illustrate the network response from a local, analytically tractable perspective, we next introduce a simple solvable model that clarifies how individual frequencies are processed and will serve as a reference point for our later analysis.}%To illustrate the response of networks as open systems to individual frequencies, we consider the following example.}

\begin{example}
{Any periodic input signal can be decomposed into different frequencies through its Fourier series decomposition, which allows the application of frequency response analysis to generic periodic input signals. We focus on the unidirectional chain network with $N$ nodes shown} in Fig.~\ref{fig:chain}, where we assume that node $1$ is the input node and node $N$ is the output node.
{For this example, the network Jacobian $A$ can be written as follows,} %For this network, the adjacency matrix $A$ is
% and the input and the output matrices $B$ and $C$ are
\begin{align}
\begin{split}
    A & = \begin{bmatrix}
     -c &  &  &  &    \\ 
     w_1 & -c &  &  &    \\ 
     &   w_2 & -c &  &    \\ 
     &  &  \ddots & \ddots &   \\ 
     &  &   & w_{N-1} & -c
    \end{bmatrix},
    % \quad B = \begin{bmatrix}
    % 1 \\ 
    % 0 \\ 
    % 0 \\ 
    % \vdots \\ 
    % 0 
    % \end{bmatrix}, \\
    % C & = \begin{bmatrix}
    % 0 & 0 & \hdots & 0 & 1 
    % \end{bmatrix}.
\end{split}
\end{align}
{where $w_i$ represents the coupling from node $i$ to node $i+1$, $i=1,...,N-1$ and $-c \leq 0$ is the stabilizing self feedback coefficient at each individual node.}
The transfer function for this network is \cite{ogata2010modern}
\begin{equation} \label{Eq:chain}
    G(s) = \dfrac{w_1 w_2 \hdots w_{N-1}}{(s+c)^N}= \dfrac{\xi_N}{(s+c)^N},
\end{equation}
{where $\xi_N=w_1 w_2 \dots w_{N-1}$ is the product of all the node-to-node couplings along the chain with $N$ nodes.} The schematics for the Bode plot of this system is shown in Fig.~\ref{fig:chain} that summarizes our results, where we sketch the Bode plot based on the parameters $c, N, w_1, \hdots, w_{N-1}$. 
{
We see that for frequencies $\omega$ below $c$, the rate of amplification is given by the DC gain, but as $\omega$ grows larger than $c$, the amplitude of the output signal decreases with a rate $\propto 1/\omega^N$. The phase plot has a peak and a low plateau corresponding to low and high frequencies, respectively. A negative phase shift arises in the output signal compared to the input signal, by as much as $-90 N$ degrees in large frequencies.
The cornering frequency is the frequency at which the power output of a system is reduced to half of its passband value.
This corresponds to a 3 dB drop in the magnitude of the signal.
Fig.~\ref{fig:chain} shows that the cornering frequency is equal to $c$.
}
\begin{figure}
    \centering
    \includegraphics[width=0.8\linewidth]{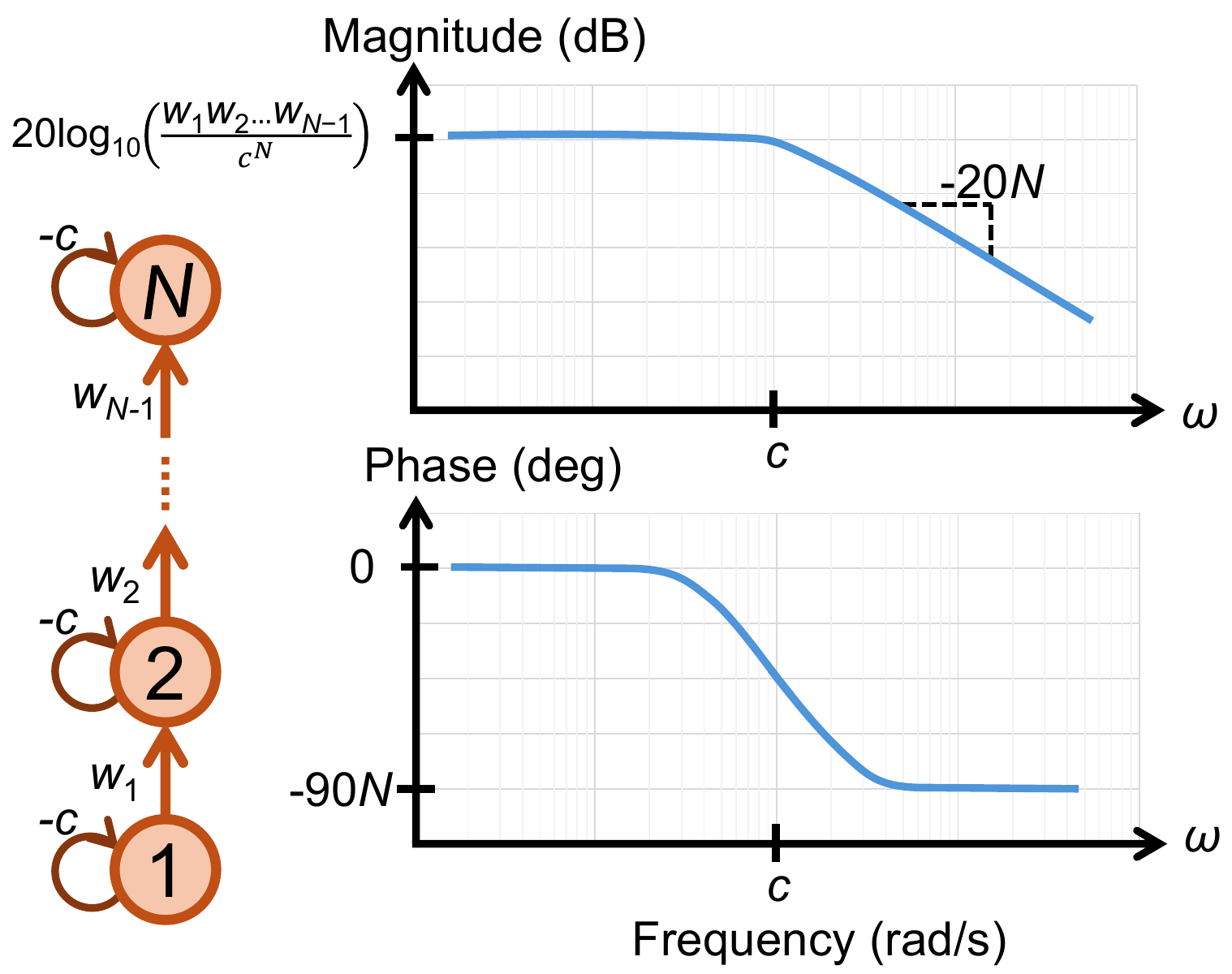}
    \caption{\textbf{Bode diagram for chain network.}
    Panel a shows the unidirectional chain network with $N$ nodes and positive link weights $w_1, w_2, \hdots, w_{N-1}$ and the self-loops with the weights $-c\leq 0$. Panel b shows the Bode plot corresponding to this system when node $1$ is the input node and node $N$ is the output node.
    In both plots, the $\omega$-axis is on a log scale.
    }
    \label{fig:chain}
\end{figure}

The $\mathcal{H}_2$-norm squared for $G$ is 
\begin{equation} \label{eq:h2normsq}
    \| G \|_2^2 =  \dfrac{\binom{2N-2}{N-1}}{2^{2N-1}} \dfrac{\xi_N^2}{c^{2N-1}}= \phi_N \dfrac{\xi_N^2}{c^{2N-1}}.
\end{equation}
The details of the calculation can be found in the Supplementary Material.
%Table \ref{table:h2} summarizes $\| G \|_2^2$ for $N = 1, 2, \hdots, 6$.

%In the Methods we study the effects of varying the length of the chain $N$ on the $\mathcal{H}_2$ norm squared in Eq.\ \eqref{Eq:chain}. 

%By introducing the geometric mean of the $w_i$'s, $\bar{w}=(w_1 w_2 \dots w_{N-1})^{\frac{1}{N-1}}$,  which represents the average node-to-node stimulation along the chain,

Table \ref{table:h2} displays $\phi_N$ for $N = 1, 2, \hdots, 6$, {from which we see that the coefficients $\phi_N$ decay sub-linearly with $N$. }

\end{example}

{
It is important to study the effects of varying the length of the chain $N$ on the $\mathcal{H}_2$ norm squared in Eq.\ \eqref{eq:h2normsq}. We first consider the case that $w_1=w_2=...=w_N=\bar{w}$, then $\xi_N=\bar{w}^{N-1}$.
For this case, we see that as the length of the chain $N$ grows, $\| G \|_2^2$ in Eq.\ \eqref{eq:h2normsq} will either increase or decrease, depending on whether $|\bar{w}|>c$ or $|\bar{w}|<c$, i.e., whether the magnitude of the node-to-node coupling stimulation  exceeds the local suppression at each node. 
Intuitively, something similar happens in the case of heterogeneous $w_i$ couplings, i.e.,  $\| G \|_2^2$  will be large (small) when $\tilde{\xi}>c$ ($\tilde{\xi}<c$), where  $\tilde{\xi}=|\xi_N|^{\frac{1}{N-1}}$ is the geometric mean of the absolute values of the node-to-node couplings along the chain with $N$ nodes.} %Namely, %$\tilde{w}=|w_1 w_2 \dots w_{N-1}|^{\frac{1}{N-1}}$.}

We also derived a closed form expression for the $\mathcal{H}_2$-norm in the case in which the self-feedback coefficients at different nodes along the chain are different $c_1\neq c_2 \neq ... \neq c_N$,
\begin{align}
    \| G \|_2^2 =  (-1)^N\xi_N^2\sum_{i=1}^{N} \dfrac{1}{2c_i}\prod_{j=1; j\neq i}^{N}\dfrac{1}{(c_i^2-c_j^2)}\,.
\end{align}
See the Supplementary Material Note 1.B for our derivations.
{ Overall, this solvable directed-chain model shows that the chain behaves as a feed-forward filter: when node-to-node stimulation exceeds local damping, signals propagate through the chain, whereas when damping dominates, the chain effectively blocks them. This pass–block behavior mirrors what occurs in more general feed-forward structures in DAG-like networks~\cite{asllani2018structure} and, as we demonstrate in the next section, also underpins signal transmission patterns in a large dataset of real-world networks.}

%the value of $\mathcal{H}_2$-norm decreases, which suggests that for larger chains, the dissipation rate $-c$ dominates the amplification rate $d$.

\begin{table}
\centering
{
\caption{
The coefficient $\phi_N$ in Eq.\,\eqref{eq:h2normsq} for different values of the system order $N$.}\label{table:h2}
\begin{tabular}{c c c c c c c}
\hline \hline
     $N$  & 1 \hspace{0.02\linewidth} & 2 \hspace{0.02\linewidth} & 3 \hspace{0.02\linewidth} & 4 \hspace{0.02\linewidth} & 5 \hspace{0.02\linewidth} & 6 \\
     \hline 
     \rule{0pt}{1.05\normalbaselineskip}
     \hspace{0.025\linewidth}
     $\phi_N$ \hspace{0.025\linewidth} & $\dfrac{1}{2}$ \hspace{0.02\linewidth} & $\dfrac{1}{4}$ \hspace{0.02\linewidth} & $\dfrac{3}{16}$ \hspace{0.02\linewidth} & $\dfrac{5}{32}$ \hspace{0.02\linewidth} & $\dfrac{35}{256}$ \hspace{0.02\linewidth} & $\dfrac{63}{512}$ \vspace{0.01\linewidth} \\
     \hline \hline
     \end{tabular} }
\end{table}

% 

% Supplementary Note 1 shows that the output standard deviation is determined by the spectrum of the output controllability Gramian, and consequently, the $\mathcal{H}_2$-norm.

% Supplementary Note 2 calculates a closed-form formula for the integral of the Bode magnitude as a function of the eigenvalues and eigenvectors of the matrix $A$.
% We also provide upper bounds for this integral using the condition number of the matrix of the eigenvectors of $A$.

% \color{black}

\subsection{Analysis of Real Networks}

It is natural to expect that many biological networks may have formed throughout evolution in order to achieve a desired balance between passing and blocking behavior. For example, certain networks may require amplifying given input signals, while others may be used to suppress disturbances and noise. At the same time, a crucial question in technological applications is how to modify an existing network to enhance its passing versus blocking characteristics.
{
In this context, both the directedness of the network—reflecting its asymmetry and hierarchical organization—and the functional distinction between input nodes (sources) and output nodes (sinks) are critical in determining signal propagation and amplification.  
Hereafter, by sources we refer to the nodes where energy, matter, or information enter the network from the environment, initiating dynamical activity that propagates through the network.  
In food webs, for instance, primary producers such as plants or phytoplankton act as sources by converting external energy (sunlight) into biomass \cite{krause2003compartments, dunne2008compilation, johnson2017looplessness};  
in power grids, generators inject electrical energy into the network \cite{athay1979practical, IEEE30, IEEE57, grigg1999ieee, 118bus, simonsen2008transient, delabays2023locating};  
in connectomes, sensory or afferent regions receive external stimuli and relay signals \cite{felleman1991distributed, young1993organization, scannell1999connectional, honey2007network, varshney2011structural, oh2014mesoscale};  
in gene regulatory networks, upstream transcription factors respond to environmental cues to regulate downstream targets \cite{xu2013escape, barah2016transcriptional, johnson2017looplessness, fang2021grndb, weinstock2024gene};  
and in signaling pathways, receptors at the cell membrane or intracellular sensors detect external signals and activate downstream cascades \cite{Laboratories_2022}.  
These examples underscore that identifying appropriate sources and sinks based on their biological or physical roles is essential for understanding and optimizing the passing or blocking characteristics of real-world networks.}

\begin{figure*}
    \centering
    \includegraphics[width=\linewidth]{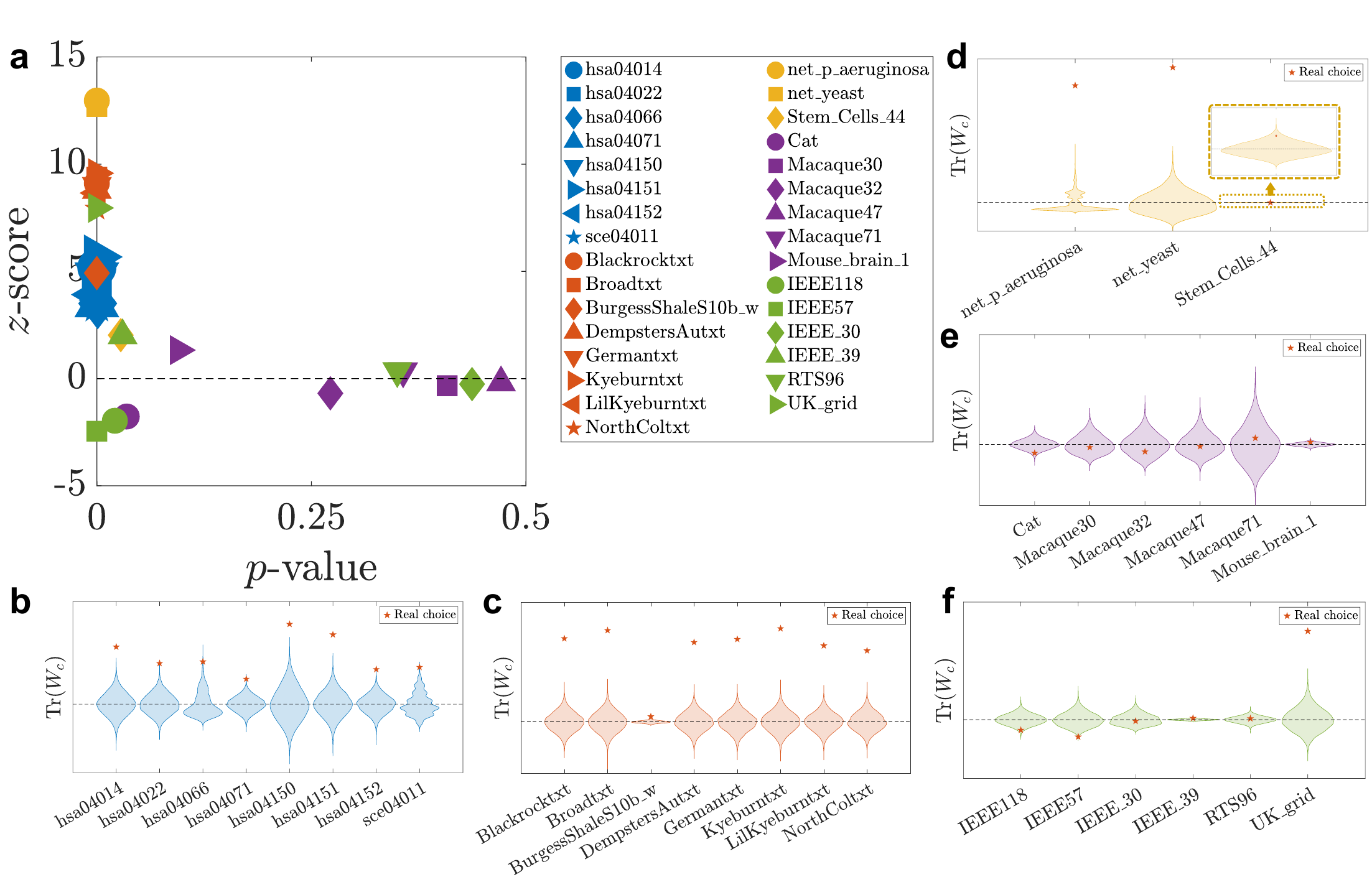}
    \caption{
    \textbf{Real data analysis.} Panel a shows the $z$-score vs $p$-value of the real data choice of the input nodes based on evaluation of the trace $\Tr (W_c)$ of real network data {(based on their Jacobian matrix $A$ and the input matrix $B$)}. Panels b-f show the distributions of $\Tr (W_c)$ for selected real networks over 10,000 sets of randomly chosen input nodes. The number of randomly input nodes is the same as the number of input nodes from the real data. For each network, the choice of real input nodes is plotted as a red star. The violin plots are shifted for better visualization, such that the mean of each distribution lies on the dashed black line. Networks within the same family are plotted using the same color in different panels: b: Pathway networks (black), c: Food Webs (red), d: Genetic network (yellow), e: Connectomes (purple), and f: Power Grids (green). }
    \label{fig:real}
\end{figure*}

%Next, we study the level of blocking versus passing behavior of several empirical networks. 
By using the analytical tools introduced earlier, particularly the $\mathcal{H}_2$-norm, we now analyze the blocking versus passing behavior of several empirical networks. 
 In this section, to accommodate different types of network dynamics, we consider the general nonlinear model introduced in Sec. A,
$\dot{z}_i(t) = f(z_i(t)) + \sigma \sum_j \tilde{A}_{ij}\, g(z_i(t),z_j(t))$,
$i=1,..,N$,  %where $z_i(t)$ is the state of system $i$,
%the function \(f\) describes the intrinsic dynamics of each node, \(g\) the pairwise interactions, \(\sigma\) the global coupling strength, and \(\tilde{A}\) the directed adjacency matrix of the network. 
where the specific forms of the functions 
$f$ and 
$g$ depend on the underlying system. These functions vary across applications—for example, they take different forms in food webs, power grids, and connectomes—as detailed in Methods, \ref{methods:models}.

%Most network datasets specify only the node-to-node connectivity (the off-diagonal entries of \(\tilde{A}\)) and provide no information about the underlying node dynamics or the entries of the Jacobian \(A\). To guarantee asymptotic stability of the linearized dynamics, we introduce a scalar shift and consider the modified operator
%\begin{equation} 
%    J \leftarrow J - c\,\mathbb{I}, \label{eq:shift}
%\end{equation}
%where \(c > \max_i \Re(\lambda_i)\) and \(\lambda_i\) are the eigenvalues of the original Jacobian \(J\). This operation shifts all eigenvalues of the Jacobian into the left half of the complex plane while leaving its eigenvectors unchanged, so that the linearized operator keeps exactly the same edges as the original network. 
}
\color{black}

{
In the analysis that follows, for each network model, we evaluate the Jacobian $A$ of the nonlinear system about a stable equilibrium, and use the pair $(A,B)$ to calculate the trace of the  Gramian $\Tr(W_c)$ and the $\mathcal{H}_2$-norm.
These Jacobian matrices may have entries with negative sign. 
For detailed information on the nonlinear models and the evaluation of the Jacobians, see the Methods \ref{methods:models}.
}
For each dataset, we fix the number of input nodes $m$ to its empirical (measured) value. 
We first choose uniformly at random 10,000 sets of $m$ nodes without repetitions, set them as input nodes, and evaluate $\Tr(W_c)$, the trace of the infinite-horizon continuous-time  Gramian.
We compare the resulting $\Tr(W_c)$ from randomly chosen input nodes with the $\Tr(W_c)$ resulting from input nodes provided by the empirical data.
The comparison is performed by evaluating two measures:
\begin{equation}
    z\text{-score} = \dfrac{x - \bar{x}}{s}, \quad p\text{-value} = \int_{x_{\min}}^{x_{real}} p(x) dx \mod 0.5,
\end{equation}
where $\bar{x}$ is the mean of the sample, $s$ is the standard deviation of the sample, $x_{\min}$ is the minimum of the sample, $x_{real}$ is the real data choice, and $p(x)$ is the probability density function such that $\int_{x_{\min}}^{x_{\max}} p(x) dx = 1$.
We define
\begin{align*}
\text{Passing network}: &
\begin{cases}
    \text{Low $p$-value}, \\
    \\
    \text{large positive $z$-score},
\end{cases} \\ 
\text{Blocking network}: &
\begin{cases}
    \text{Low $p$-value}, \\
    \\
    \text{large negative $z$-score}\,,
\end{cases}
\end{align*}
{and we recall that extreme (large magnitude) \( z \)-scores correspond to low \( p \)-values, indicating statistically \textit{atypical} configurations, while small \( z \)-scores and high \( p \)-values reflect \textit{typical} behavior.} { In simple terms, large positive $z$-scores correspond to networks that predominantly \emph{pass} (amplify) inputs, large negative $z$-scores correspond to networks that predominantly \emph{block} (attenuate) them, and values near zero indicate networks that are neither particularly passing nor particularly blocking.}
The results of $z$-score vs $p$-value are shown in Figure~\ref{fig:real} a. We consider different categories of networks: power grids for which input nodes are generators,  connectomes for which input nodes are sensory neurons, 
molecular signaling networks for which input nodes are receptors, gene regulatory networks for which input nodes are transcription factors, pathway networks for which input nodes are extracellular ligands, and food webs for which input nodes are autotrophs (plants and algae) that produce their food via photosynthesis.
{
For a detailed explanation of the real network dataset, see Supplementary Note 4.
}
For all networks, in the absence of more detailed information, we set $C=I$.
{
The violin plots in Fig.~\ref{fig:real} b-c show the distribution of $\Tr (W_c)$ along with the real network choice. 
These distributions and the real network choice are used to evaluate the $z$-score and $p$-value shown in panel a.
}

Figure~\ref{fig:real} shows that in pathway networks, food webs, and genetic networks, the real choice of input nodes results in passing behavior, while power grids and connectomes do not show a discernible pattern, since some networks are passing, some are blocking, and the rest are typical.
The case of power grids is interesting as some of these networks are the most blocking. There are two reasons for this, the first one being that in these networks, input nodes are the generator buses where the power is injected into the grid, and the power flows on the transmission lines are given by the voltage angle differences. Typically, the generator buses either have many transmission lines to their neighbors, or a few lines with a high capacity, allowing the power flow to circulate with small angle differences and therefore ensuring the stability of the grid. One thus expects a power grid with well connected generator to be blocking in general. The second reason is that, at odds with the other types of networks considered here, power grids have a Jacobian matrix $A$ that is close to symmetric. The latter corresponds to the Jacobian of the voltage angle dynamics close to the operational state. In a grid where the resistance is negligible compared to the reactance, $A$ is symmetric. Inclusion of the transmission line losses adds a slight asymmetry to $A$, which remains close to a symmetric matrix.
The violin plots confirm our conclusions while providing a more detailed view, as they display the entire distribution: atypical values appear in the tails, whereas typical values cluster near the peak, usually centered around the mean. {Connectomes represent another class of networks characterized by notably stronger blocking behavior. In Subsection D we explain this in terms of their weak directionality and non-normal properties.}
\color{black}

{
Complementary results are reported in Supplementary Notes~5, where adjacency matrices of the real data is used instead of the Jacobian matrix in the evaluation of the $\mathcal{H}_2$-norm, yielding comparable results.
}
In Supplementary Note~6, we consider the largest eigenvalue of the Gramian, \(\lambda_{\max}(W_c)\), as a measure of the maximum amplification rate achievable by input signals. The findings are analogous to those reported in the main text, confirming the same qualitative distinctions between passing and blocking regimes across network types. Supplementary Note~7 further presents histograms of the \(\mathcal{H}_2\)-norm for empirical networks, comparing two scenarios: (i) the case in which all nodes serve as input nodes, and (ii) the case in which 
$m$ input nodes are selected at random, with 
$m$ equal to the number of inputs in the empirical dataset.

% 

% Similar analysis of empirical networks in Fig.\,\ref{fig:real} is performed in Supplementary Note 4, where instead of $\Tr (W_c)$, the largest eigenvalue of the controllability Gramian $\lambda_{\max} (W_c)$ is the measure of interest.
% We show that $\lambda_{\max} (W_c)$ denotes the maximum amplification rate that a series of input signals may experience.
% Our analysis of real network data in this case reveals that Pathway networks, Food webs, and Genetic networks, the real data choice of the input nodes, results in a passing behavior.
% Power Grids tend to show a mix of blocking and typical behavior, while Connectomes do not show a discernible pattern since some networks are passing, some are blocking, and the rest are typical.

% Supplementary Note 5 provides histogram plots of the $\mathcal{H}_2$-norm of selected real network datasets where (i) all nodes are input nodes and (2) $m$ number of nodes are randomly chosen as input nodes, where $m$ is the number of input nodes in each real network dataset.

% \color{black}

% Next, we provide intuition of why the values of the trace of the Gramian (or equivalently $\mathcal{H}_2$-norm squared) are different for different classes of networks.

\subsection{Directedness as a Passing Mechanism}

Systems such as food webs, signaling pathways, and gene regulatory networks tend to exhibit larger \( z \)-scores and stronger directed hierarchical structure, being almost or strictly DAGs, consistent with their high non-normality and directional flow~\cite{o2021hierarchical}. Non-normality, which strongly impacts the behavior of dynamical systems~\cite{asllani2018topological, MUOLO2019Patterns, nazerian2023Communications}, is a prominent feature in many naturally occurring networks, as shown in Refs.~\cite{asllani2018structure,o2021hierarchical}.
In contrast, power grids and neural networks are either blocking or typical. %, and neuronal networks do not show a discernible pattern since some networks are passing, some are blocking, and the rest are typical. 
This is consistent with the fact that power grids tend to be symmetric, in general, and connectomes exhibit weak directedness or non-normality (see also Ref.~\cite{o2021hierarchical}). Based on these structural properties, one can conjecture that most real-world networks, due to their inherent directedness, are expected to exhibit passing behavior under optimal input selection. To shed light on this conjecture, we next examine why the trace of the Gramian (or equivalently the squared $\mathcal{H}_2$-norm) varies across network classes, focusing on their directedness as captured by %the non-normality measured through 
the Henrici index \cite{higham2005spectra}.

%We analyze the level of non-normality of the connectivity matrix among different groups of real networks.
{
The normalized Henrici index $\hat{d}_F(A)$ is a measure of departure from normality \cite{asllani2018structure} for a  matrix $A$ based on the Frobenius norm and is defined as
\begin{equation} \label{eq:henrici}
    \hat{d}_F(A) := \dfrac{\sqrt{\| A\|_F^2 - \sum_{i=1}^N | \lambda_i |^2}}{\| A \|_F}.
\end{equation}
Here, $\| \cdot \|_F$ denotes the Frobenius norm, and $\lambda_i$ are the (possibly) complex eigenvalues of the matrix $A$. 
If the matrix $A$ is normal, then $\hat{d}_F(A) = 0$.
If the matrix $A$ is triangular (highest structural non-normality), then $\hat{d}_F(A) = 1$.
}

% Figure~\ref{fig:henrici} shows the normalized trace of the controllability Gramian $\Tr(W_c) / N^2$ vs the Henrici index $\hat{d_F}(J)$.
% We see a general trend that the higher the Henrici index, the higher the normalized trace.
% Also, networks within the same category appear to be clustered in this plot, showing that different network categories correspond to different levels of non-normality and of the normalized trace.

Figure~\ref{fig:henrici} shows the normalized trace of the  Gramian, $\Tr(W_c) / N^2$, versus 

the Henrici index $\hat{d}_F(A)$ of the Jacobian matrix $A$. The data is generated in the same way as in Fig. \ref{fig:real}.
We observe a clear trend: higher Henrici index correlates with higher normalized trace. Moreover, networks in the same category cluster together in this plot, revealing distinct levels of non-normality and signal amplification across categories. Notably, food webs and signaling pathways exhibit a pronounced scale separation—one to two orders of magnitude higher in normalized trace compared to other networks. This is consistent with their (almost) perfect directed acyclic graph (DAG) structure, which maximizes passing behavior along input–output paths. Interestingly, while food webs and pathways achieve this via fewer but longer paths, gene regulatory networks exhibit many very short paths, resembling a collection of directed star graphs \cite{o2021hierarchical}. 
As further detailed in the Supplementary Note 4, the normalized trace $\Tr(W_c) / N^2$ and the unnormalized $\Tr(W_c)$, both evaluated using the adjacency matrices $\tilde{A}$, in place of Jacobians, reinforce this observation by clustering gene regulatory networks closer to food webs and pathways, underscoring the role of perfect directedness in enhancing the passing property. 
\color{black}
We will explore these structural mechanisms further in the next section.

\begin{figure}
    \centering
    \includegraphics[width=0.8\linewidth]{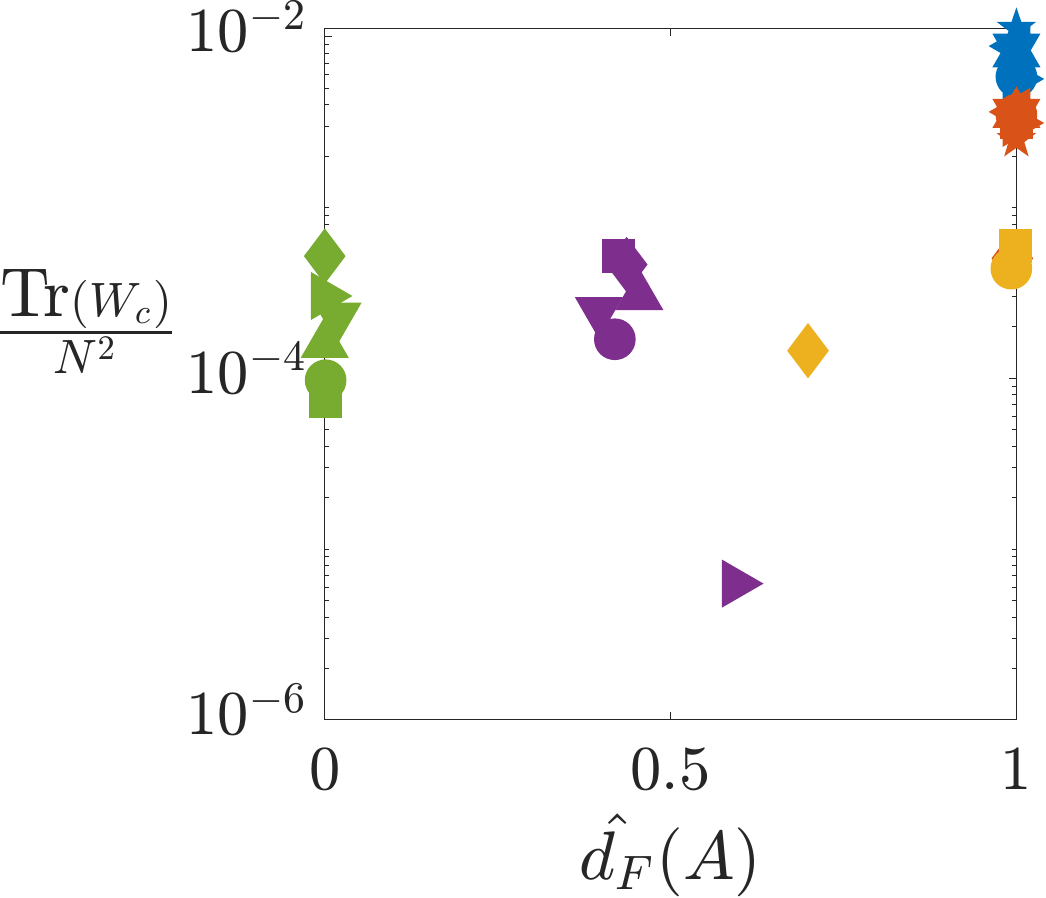}
    \caption{\textbf{Normalized trace of Gramian vs. the non-normality index.} 
    {
    The normalized Henrici index $\hat{d_F}(A)$ of the Jacobian matrix $A$ is defined in Eq.\,\eqref{eq:henrici}. 
    }
    The data is displayed using the same color and marker conventions as those in Fig.\,\ref{fig:real}. Pathway, Food web, Genetic, Connectome, and Power Grid networks are in black, red, yellow, purple, and green color, respectively. }
    \label{fig:henrici}
\end{figure}

\color{black}

\subsection{Directed Acyclic Graphs}

{ Building on the previous findings that highlight the importance of directed acyclic graph (DAG) structures—both for their prevalence in real-world networks and their role in facilitating signal propagation—this section adopts a complementary approach to gain analytical insight into how the shape of directed paths influences the amplification of signals along input–output pathways.}

{We focus on open networks defined by the triplet \( (A,B,C) \). The off-diagonal entries of the matrix $A$ represent the connectivity of a DAG, while the diagonal entries are negative quantities that represent the stabilizing self
feedback coefficient at each individual node.}
%Here, \( P = [P_{ij}] \), with \( P_{ij} \geq 0 \), represents the adjacency matrix of a DAG, and \( D \) is a diagonal matrix with non-negative entries. 
{By construction, here, we take the matrix \( A \) to be a Metzler matrix, {i.e., a matrix for which all the off-diagonal entries are nonnegative}}. The quantity of interest is the \( ji \)-th entry of \( A^{-1} \), which corresponds to the DC gain when \( i \) is chosen as the input node and \( j \) as the output node.
}\color{black}
Following \cite[Theorem 6.4]{briat2017sign}, it is easy to calculate
\begin{equation} \label{eq:dag0}
    [A^{-1}]_{ji} = \sum_{\text{directed path } \pi: i \to j} \left( \prod_{k \to l \in \pi}  \dfrac{A_{lk}}{-A_{kk}} \right) \dfrac{1}{A_{jj}}.
\end{equation}
The term $A_{lk} / (-A_{kk})$ inside the product denotes the edge gain divided by the local leak. 
Ultimately, the contribution of the gains and leaks along the path is divided by the local leak at the output node $j$.
Since we require the matrix $A$ to have all eigenvalues with negative real parts for stability, it means the main diagonal of the matrix $A$ has all strictly negative entries, i.e., $A_{ii} <0$, $\forall i$. Eq.\,\eqref{eq:dag0} can be simplified to
\begin{equation} \label{eq:dag}
    [A^{-1}]_{ji} = -\sum_{\text{directed path } \pi: i \to j} \left( \prod_{k \to l \in \pi}  \dfrac{A_{lk}}{|A_{kk}|} \right) \dfrac{1}{|A_{jj}|}.
\end{equation}

{It is important to highlight the intuitive interpretation of Eq.~\eqref{eq:dag}.  
%This observation is crucial because it shows
We see that the passing, amplifying, or damping behavior of DAG networks as open systems  
depends both on the number of directed paths from input to output and on the weights of the edges that make up these paths.  
Each edge weight contributes to amplification if it is larger than one, or to attenuation if it is smaller than one;  
 longer paths compound these effects. Moreover, the self-loops always contribute to damping—since in our formulation  
they correspond to the local stability term \( c \)—and appear in the denominator of each term in the sum.  
Based on this reasoning, we can also understand why gene networks, which have many but very short (often one-step) paths,  
exhibit a considerable decrease in the normalized trace \( \Tr(W_c) / N^2 \), whereas food webs and signaling pathways,  
with fewer but longer weighted paths, maintain higher normalized trace values.}

{In conclusion, the DC gain for a directed acyclic graph, can then be expressed as
\begin{equation}
    \text{DC Gain} = 20 \log_{10} \left| [A^{-1}]_{ji} \right|,
\end{equation}
where \( [A^{-1}]_{ji} \) is given explicitly in Eq.~\eqref{eq:dag}.} Next we present an illustrative example.
%Intuitively, we can rewrite \eqref{eq:dag} as the sum of fractions that have ``Product of the weights along the path" as numerators and ``Product of absolute value of self-loop weights of visited nodes" as denominators.

% In the case of directed acyclic graphs, the DC gain becomes
% \begin{equation}
%     \text{DC Gain} = 20 \log_{10} \left| [A^{-1}]_{ji} \right|,
% \end{equation}
% where $[A^{-1}]_{ji}$ is given in Eq.\,\eqref{eq:dag}. 

% Intuitively, we can rewrite \eqref{eq:dag} as
% \begin{equation}
%     [A^{-1}]_{ji} = -\sum_{\text{directed paths } i \to j}  \dfrac{\text{Product of the weights along the path}}{\text{Product of absolute value of self-loop weights of visited nodes}}.
% \end{equation}

\begin{example}
Figure\, \ref{fig:dag} shows a directed acyclic graph (without considering self-loops) and its {associated} %adjacency 
matrix $A$. 
For this network, node $i=1$ is the input node and node $j=5$ is the output node.
There are $3$ paths from node 1 to 5: $\pi_1=( 1,2,4,5 )$, $\pi_2=( 1,2,3,4,5 )$, and $\pi_3=( 1,3,4,5 )$.
Therefore, applying Eq.\,\eqref{eq:dag} results in
\begin{align*}
    [A^{-1}]_{51}  = & -\dfrac{A_{21} A_{42} A_{54}}{A_{11} A_{22} A_{44} A_{55}} + \dfrac{A_{21}A_{32} A_{43} A_{54}}{A_{11} A_{22} A_{33} A_{44} A_{55}} \\
    & - \dfrac{A_{31} A_{43} A_{54}}{A_{11} A_{33} A_{44} A_{55}}.    
\end{align*}
Note that since we assume negative self-loops for stability, then
\begin{align*}
    [A^{-1}]_{51} = & -\dfrac{A_{21} A_{42} A_{54}}{|A_{11} A_{22} A_{44} A_{55}|} - \dfrac{A_{21}A_{32} A_{43} A_{54}}{|A_{11} A_{22} A_{33} A_{44} A_{55}|} \\
    & - \dfrac{A_{31} A_{43} A_{54}}{|A_{11} A_{33} A_{44} A_{55}|}.
\end{align*}

\begin{figure}
    \centering
    \includegraphics[width=\linewidth]{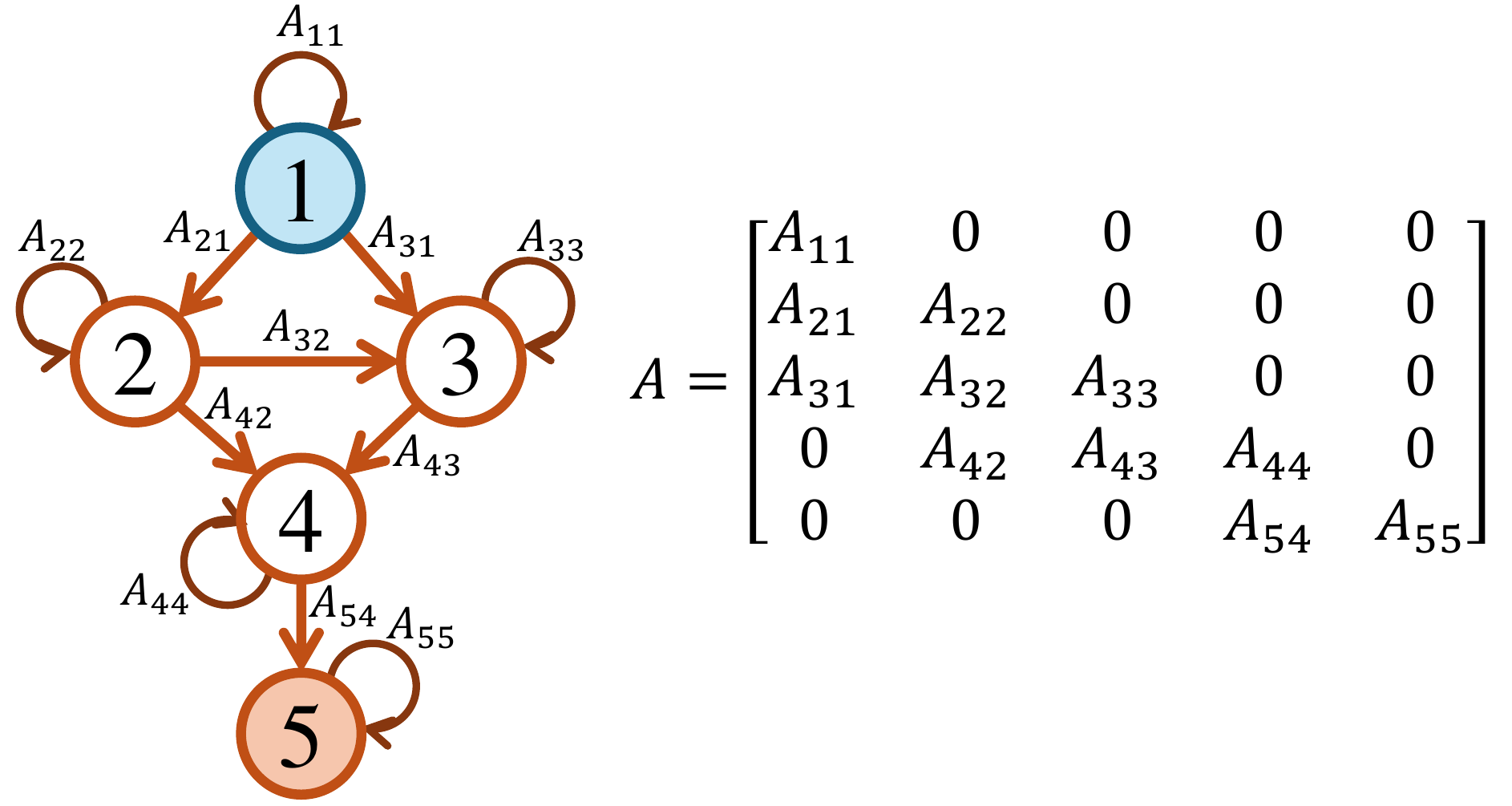}
    \caption{
    \textbf{Directed Acyclic Graph (excluding self-loops) and its corresponding Jacobian matrix $A$.} The input and the output nodes are shown in black and orange, respectively. Self-loops are considered to provide asymptotic stability of the linear dynamics.}
    \label{fig:dag}
\end{figure}
\end{example}

\subsection{Frequency Response of General Networks}

\begin{figure}
    \centering
    \includegraphics[width=0.8\linewidth]{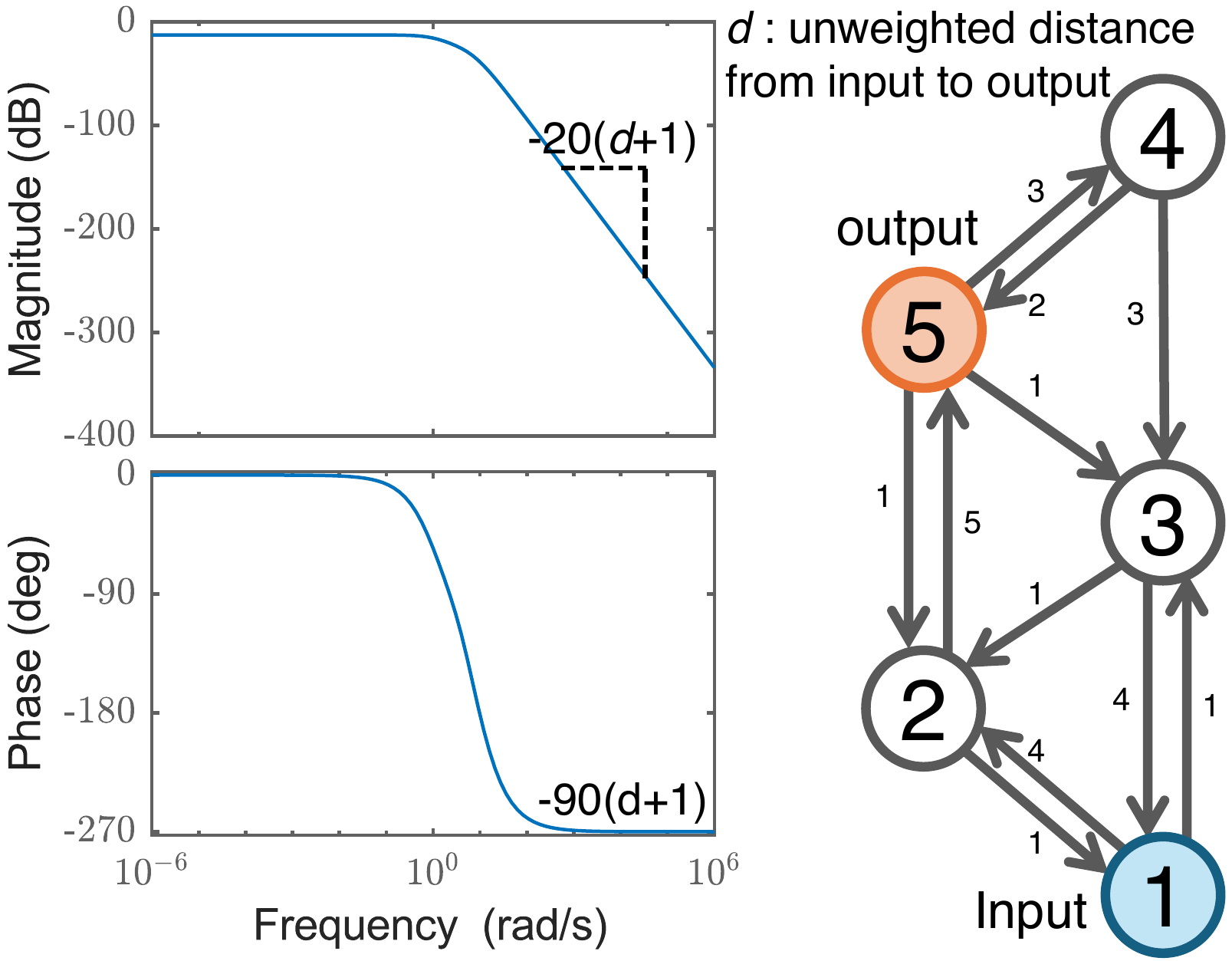}
    \caption{\textbf{Bode diagram for a general weighted network.}
    Bode plot (left panels) of the network shown on the right. The matrix $A$ defining the topology of the network is shifted as $A \leftarrow A - c I$ such that the largest real part eigenvalue of the matrix $A$ is $-1$.
    All nodes have self-loops of weight $-c$, which are not shown for better visualization.
    The shortest path without considering edge weights is from the input node to node 2 and then to the output node, which has a length of $d = 2$.
    The slope of the magnitude after the cornering frequency $c = 1$ is slope $=-60$, and the final phase $=-270$.}
    \label{fig:bode_example}
\end{figure}

{Next, we consider general network topologies—without requiring a DAG structure—while retaining the assumption that the matrix 
$A$ is Metzler and has negative real part eigenvalues.} 
We find that the slope of the magnitude plot {for large frequencies ($\omega \gg$ cornering frequency)} 
and the final phase for large frequencies follows the general relations,
\begin{equation}
    \text{slope} = -20(d+1) , \qquad \text{final phase} = -90(d+1), 
\end{equation}
where $d$ is the shortest path length from the input node to the output node without considering the weights.
Alternatively, $d+1$ is the effective order of the dominant pole at the cornering frequency $c$, where $-c$ is the largest real-part eigenvalue of the matrix $A$.
Figure~\ref{fig:bode_example} confirms our above relations for the slope, the final phase, and the cornering frequency.
Note that in this example, if one considers the weights when evaluating $d$, then the path from the input node to node 2 to the output node is no longer the shortest, while the shortest path involves visiting node 4 before node 2.
The evaluation of the weighted shortest path does not match the numerical results shown in the left panels of Fig.~\ref{fig:bode_example}, while our definition of the shortest unweighted path length matches.

\section{Conclusion}

In this work, we  propose a general theory to study how networked dynamical systems behave as open systems—processing, amplifying, or attenuating environmental inputs depending on their internal structure and the specific configuration of input and output nodes. We employ the $\mathcal{H}_2$-norm as a robust and universal metric to quantify the degree of input-to-output signal amplification across a wide class of inputs, including sinusoidal, periodic, and stochastic signals. { Within this framework, we show how the frequency response of an arbitrary network, including its Bode profile, is affected jointly by the underlying topology and by the choice of input and output nodes. We also examine the $\mathcal{H}_2$-norm across several empirical networks and derive new closed-form results for directed acyclic graphs and for the zero-frequency limit. In addition, we obtain an asymptotic approximation for the trace of the  Gramian in the regime dominated by strongly stable dynamics, and we uncover a modularity property of the squared $\mathcal{H}_2$-norm that applies to both input and output nodes, extending previous results that focused only on inputs.}

%{ The new contributions of this work include: (i) our analysis of the frequency response for an arbitrary network and our sketch of the Bode plots in terms of the network topology and the particular selection of the input and output nodes, (ii) the study of the $\mathcal{H}_2$-norm for several empirical networks, (iii) our derivations for the case of directed acyclic graphs, (iv) our derivations for the DC gain, (v) our approximate formula for the trace of the Gramian in the case in which the largest real part of the eigenvalues of the matrix $A$ is large in magnitude (Eq.\ 20) and (vi) the modularity property for the $\mathcal{H}_2$-norm squared (to the best of our knowledge, the modularity was known with respect to the input nodes but not with respect to the output nodes.)} 
% {\textbf{Potential edits if only continuous:} Our formulation applies to both continuous-time and discrete-time systems.} % and connects signal response properties to the topology of the underlying network and its controllability Gramian.

Our theoretical results demonstrate that the signal amplification characteristics of a network are strongly influenced by structural factors such as the location of input and output nodes, the node-to-node weighted connectivity, and unweighted shortest path distances. We establish relationships between the network topology and signal transmission behavior, providing conditions under which a network tends to either pass or block signals, based on frequency response and DC gain analyses. Importantly, we identify {network topologies that} exhibit optimized configurations, either facilitating transmission (passing) or suppressing disturbances (blocking), reflecting functional demands of the network with its environment and constraints arising from evolution or design. {The work in \cite{nazerian2025open} investigates open networks in discrete time; however, it does not consider their frequency response.}

% Our work led to the formulation of efficient computational tools, such as the network index $\alpha$, to evaluate the influence of input node placement without computing the full Gramian matrix. 
An empirical analysis {of networks} across diverse domains, including power grids, gene regulatory systems, and neural connectomes, supports our theoretical predictions and highlights the broad applicability of our framework. We find that food webs, signaling pathways, and gene regulatory circuits are often structurally organized to enhance the passing of signals,  while the structure of engineered systems like power grids tends to suppress signal and noise propagation. Overall, this study lays the groundwork for systematic characterization and optimization of open networks, with potential applications in network design, control, and the interpretation of naturally occurring network dynamics.
{This exploration of empirical network structures, together with the dynamics they are supporting complement and further previous works which focused on the effect of topology and nonlinearities on the propagation of perturbations or on specific network structures that amplify or reduce fluctuations~\cite{hens2019spatiotemporal,meena2023emergent,tyloo2022layered}.}

Most of previous effort in investigating the response of network-coupled dynamical systems to external inputs has focused on undirected networks. 
%The response of non-symmetrically coupled, i.e., directed networked systems, has been much less studied. 
%Directed networks are typically more challenging to investigate %, due to the non-symmetric coupling. 
Many of the results obtained for undirected networks do not generalize to directed ones, which are typically more challenging to investigate.
In this work, we have applied tools and obtained results that apply to generic (non-normal) network topologies and uncovered that directedness can be used as a fundamental passing mechanism. {In particular, our data analysis reveals that directed acyclic graphs (DAGs)—a topological feature that is prevalent in food webs, signaling pathways, and gene regulatory networks—exhibit strong passing behavior. This empirical observation aligns with our analytical results for DAGs, which demonstrate that amplification and attenuation depend explicitly on the number and length of the weighted input–output paths, as well as on attenuation effects at the node level.} { Taken together, these findings are consistent with the previous studies of directed, non-normal networks, where strong asymmetries in the interaction pattern generate source–sink hierarchies, suppress back-propagation, and bias flows directionality. In such architectures, which are often structurally close to acyclic and possess a dominant directed backbone with only weak feedback loops, perturbations are funneled toward terminal components and sink nodes, naturally favoring one-way signal transfer~\cite{johnson2017looplessness,asllani2018structure,asllani2018topological,o2021hierarchical,ramon2024entropy}.}

{ While our analysis captures how network structure shapes signal propagation, it relies on the linear dynamics encoded in the Jacobian of the underlying nonlinear system. As we now emphasize, the LTI description reflects only the first–order behavior around a fixed point, and different nonlinear models defined on the same structure—or even different equilibria of the same model—can yield different Jacobians and thus different response patterns. This is an inherent limitation of any linearization-based approach. Nevertheless, linear analysis often provides meaningful insight into regimes where systems operate near criticality or in weakly nonlinear phases, where the Jacobian accurately captures local signal propagation. Outside these conditions, fully nonlinear effects become essential and lie beyond the scope of the present work. \textcolor{black}{We also note that some of our results,  in particular those in Secs.\ II E and II F, are limited to the case of non-negative weights associated with the network edges.}}

Our work provides direct insight into how to tune the $\mathcal{H}_2$-norm of an open network—either increasing it by strategically adding input or output nodes, or reducing it by removing them. This insight enables a principled approach to ranking nodes based on their contribution to the network’s amplification properties, leveraging the modularity of the output  Gramian. These findings have direct implications for engineering and bioengineering applications, for which achieving an optimal balance between blocking and passing behavior is often essential.

 {Several key questions about how complex systems respond to perturbations remain open. For instance, time-varying coupling—arising from failures or from the intrinsic evolution of the underlying network—may strongly affect signal transmission, and the impact of control actions triggered by propagating perturbations is still poorly understood. Furthermore, mounting evidence points to the pivotal role of higher-order (beyond pairwise) interactions in networked systems~\cite{PhysRevE.101.022308,majhi2022dynamics,della2023emergence,carletti2023global,bick2023higher,millan2025topology}, raising the question of how such multi-body couplings modulate their response and resilience to perturbations. Addressing these issues represents an important direction for future work.}

% Our work provides direct insight on how to either maximize the $\mathcal{H}_2$-norm of an open network by either adding input or output nodes, or to minimize the $\mathcal{H}_2$-norm of a network by removing input or output nodes. This has direct implications in engineering application, where achieving a desired level of blocking versus passing behavior may be desired.
% {Amir, please explain why this becomes a trivial task based on our study.} {Done!}
% \textcolor{teal}{
% The selection of a set of optimal input or output nodes has become a trivial scoring mechanism of each node based on the modularity property of the trace of the output controlability Gramian, which is equal to $\mathcal{H}_2$-norm squared.
% }

% An important observation is that while many networks may be designed to achieve a certain overall level of passing versus blocking behavior, in practice they possess specific capabilities to pass certain signals and block others. This is also explained by our analysis to characterize the $\mathcal{H}_2$-norm of individual input node-output node pairs, which enables direct computation of particular amplification over relevant pathways.

\section{Methods}

\subsection{From Nonlinear to Linearized equations} \label{methods:A}

We start from the general set of equations for the network dynamics we consider in this paper,
\begin{equation}
    \dot{z}_i(t) = f(z_i(t)) + \sigma \sum_j \tilde{A}_{ij}\, g(z_i(t),z_j(t)), \label{eq:orig}
\end{equation}
$i=1,..,N$, where $z_i(t) \in \mathbb{R}$ is the state of system $i$,
the function \(f : \mathbb R \rightarrow \mathbb{R}\) describes the intrinsic dynamics of each node, \(g : \mathbb R \times  \mathbb R \rightarrow \mathbb{R}\) the pairwise interactions, the scalar \(\sigma\) the global coupling strength, and \(\tilde{A} \in \mathbb{R}^{N \times N} \) the directed adjacency matrix of the network. The specific forms of the functions 
$f$ and $g$ depend on the underlying network dynamics—for example, they differ for food webs, power grids, and connectomes—as detailed in the Methods Section \ref{methods:models}.

Linearization of Eqs.\ \eqref{eq:orig} around a stable fixed point \(z_i^*\), $i=1,...,N$, yields the LTI dynamics,
\begin{equation}
    \dot{x}_i(t) = A_{ii}\, x_i(t) + \sum_{j\neq i} A_{ij}\, x_j(t), \label{eq:linear}
\end{equation}
$i=1,...,N$, with
\(A_{ii} = f'(z_i^*) + \sigma \sum_j \tilde{A}_{ij}\,\partial_1 g(z_i^*,z_j^*) \)
capturing the local stability of the intrinsic dynamics together with the self-dependence of the interactions, and
\(A_{ij} = \sigma \tilde{A}_{ij}\,\partial_2 g(z_i^*,z_j^*)\)
encoding the linearized influence of node \(j\) on node \(i\). Note that this equation is equivalent to our main equation \eqref{eq:continuous} in the absence of inputs $u(t)=0$.

As an example, we consider an $N=6$-node network with adjacency matrix,
\[
\tilde{A} = \begin{bmatrix}
0 & 1 & 0 & 0 & 0 & 0 \\
0 & 0 & 1 & 1 & 0 & 0 \\
0 & 0 & 0 & 0 & 1 & 0 \\
0 & 0 & 1 & 0 & 1 & 0 \\
0 & 0 & 0 & 0 & 0 & 1 \\
0 & 0 & 1 & 0 & 0 & 0
\end{bmatrix}.
\]
With this choice of the adjacency matrix, we adopt the Michaelis-Menten model \cite{karlebach2008modelling}, a standard framework for describing the dynamics of gene-regulatory and metabolic networks,
\begin{equation}
    \dfrac{d z_i (t)}{dt} = -\alpha_i z_i^a (t) + \kappa \sum_{j=1}^N \tilde{A}_{ij}  \dfrac{z_j^h (t)}{1+z_j^h (t)}, 
\end{equation}
$i = 1,\hdots, N$, where $z_i (t)$ represents the expression of gene $i$ at time $t$ and wthout loss of generality, we set $a=2$, $h=1$, $\kappa = 1$, and $\alpha_i=1$ at all nodes. Further details on this model can be found in the Methods \ref{methods:models}. 

We compute the Jacobian about the fixed point $\bz^*=[ 0.692,    0.921,    0.618,    0.874,    0.618,    0.618]^T$,

\[
{A} = \begin{bmatrix}
-1.384 & 0.271 & 0 & 0 & 0 & 0 \\
0 & -1.842 & 0.382 & 0.2847 & 0 & 0 \\
0 & 0 & -1.236 & 0 & 0.382 & 0 \\
0 & 0 & 0.382 & -1.748 & 0.382 & 0 \\
0 & 0 & 0 & 0 & -1.236 & 0.382 \\
0 & 0 & 0.382 & 0 & 0 & -1.236
\end{bmatrix}.
\]

We note the similarity in the structure of the matrices $A$ and $\tilde{A}$ above. In particular, the Jacobian $A$ corresponds to a weighted version of the network adjacency matrix $\tilde{A}$ plus a diagonal matrix, since the off-diagonal entries of the Jacobian multiply the original entries of the adjacency matrix, and self loops might exist. When the fixed point is homogeneous across nodes, these weights reduce to a uniform rescaling of the adjacency matrix. Application of a spectral shift $A \leftarrow A -c I$, $c>0$,  only adds a uniform damping term and does not affect the eigenvectors of the matrix $A$. As a result, the network structure and qualitative dynamical response are preserved, with the shift modifying only the global timescale~\cite{yan2015spectrum,gao2014target,klickstein2017energy}.                    
%if $r_i > 0$, or death if $r_i < 0$. The positive constants $s_i$ represent the finite carrying capacity of the ecosystem (limited resources) and prevent species $i$ from growing indefinitely.The interaction among species is given by the adjacency matrix $\tilde{A}=[\tilde{A}_{ij}]$, which has zeros on its main diagonal, i.e., $\tilde{A}_{ii}=0, \forall i$.

\subsection{Frequency Response Example} \label{methods:example}

\begin{figure}
    \centering
    \includegraphics[width=0.95\linewidth]{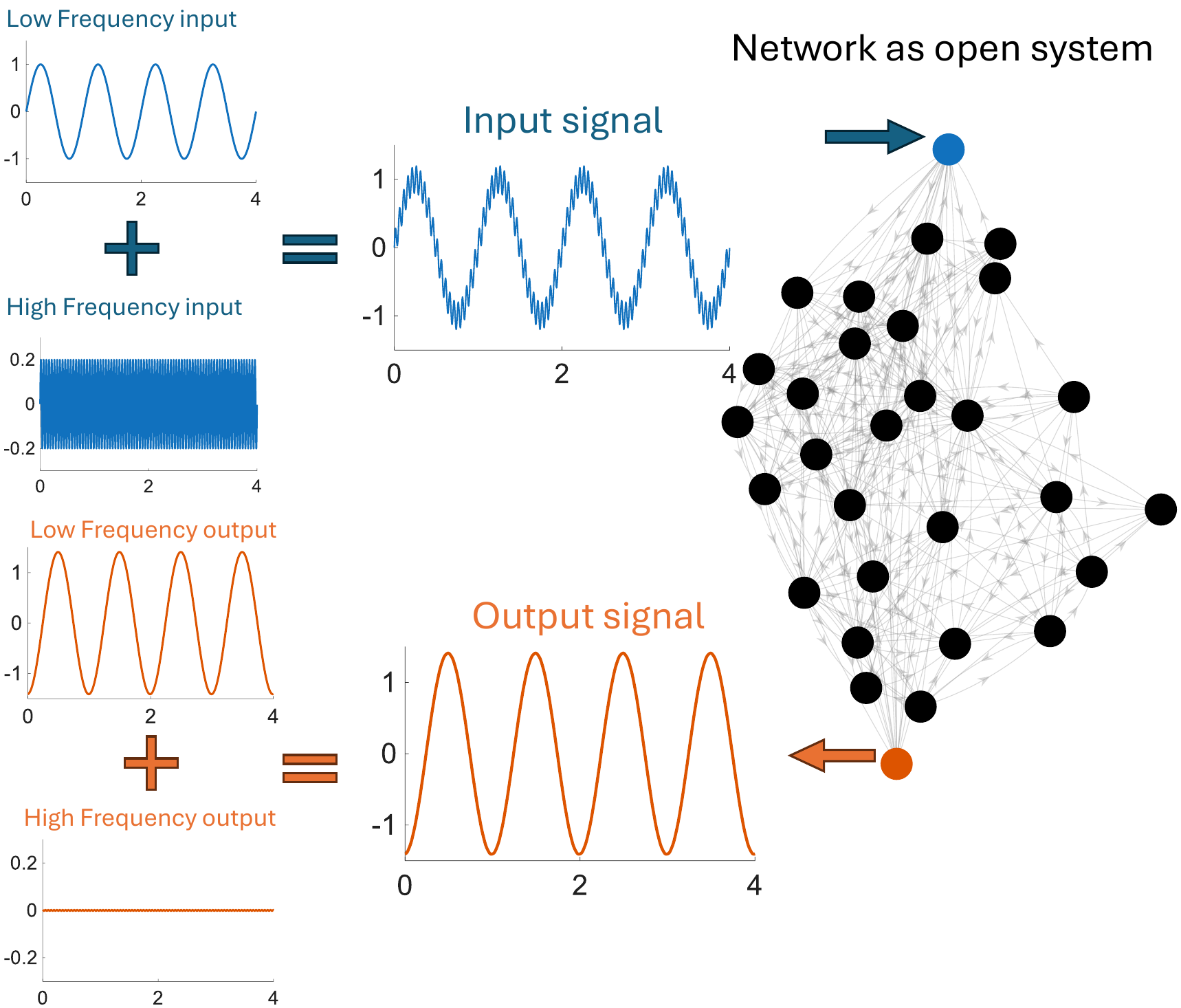} \\
    \vspace{0.02\linewidth} 
    \includegraphics[width=0.95\linewidth]{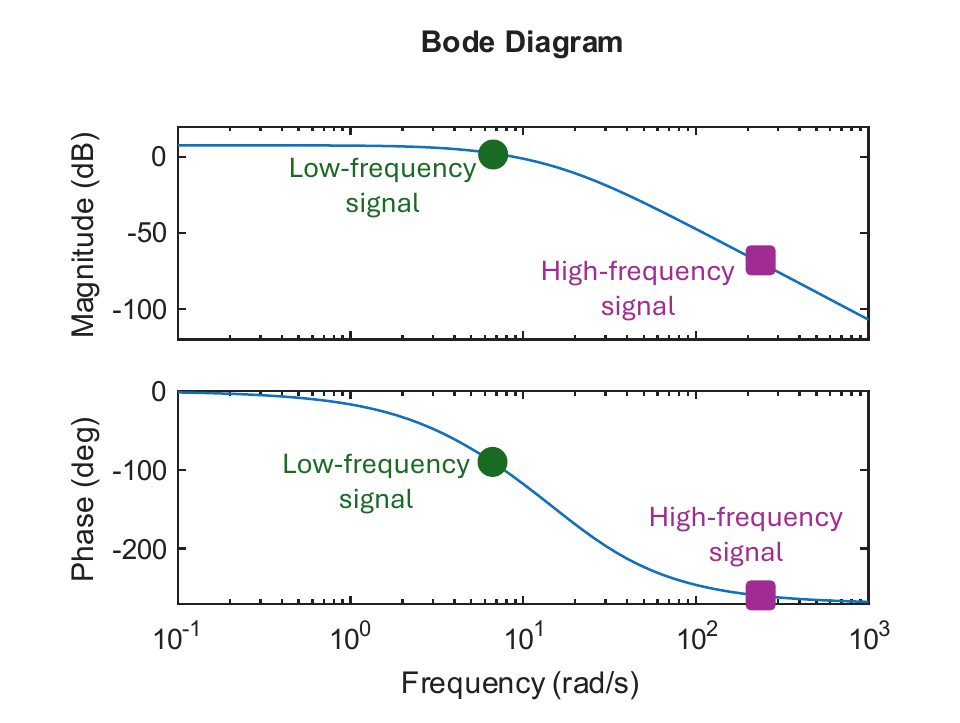}
    \caption{
    The top panel shows the passing behavior of a low-frequency input signal and blocking a high-frequency one. 
    The bottom panel shows the Bode plot associated with the system.}
    \label{fig:signal}
\end{figure}

Figure \ref{fig:signal} shows an input signal that is composed of two sinusoidal signals with two different frequencies and two different amplitudes.
The input signal enters the network through the black node and results in the output signal through the red node.
The output signal has the same two frequencies as the input signal, but with different amplitudes and phases.
We see that the amplitude of the low-frequency input is slightly amplified (from $1 \to 1.4$), while the amplitude of the high-frequency input is damped (from $0.2 \to  0.002$).
\\

\textbf{Technical details:}
\\
% The triplet $(A,B,C)$ are:
% \begin{equation}
%     A = \left[\begin{array}{ccccc} -10.22 & 1 & 4 & 0 & 0\\ 4 & -10.22 & 1 & 0 & 1\\ 1 & 0 & -10.22 & 3 & 1\\ 0 & 0 & 0 & -10.22 & 3 \\ 0 & 5 & 0 & 2 & -10.22 \end{array}\right], \quad 
%     B = \left[ \begin{array}{c} 10\\ 0\\ 0\\ 0\\ 0 \end{array} \right], \quad C = \left[\begin{array}{ccccc} 0 & 0 & 0 & 0 & 10 \end{array}\right].
% \end{equation}
The pair $(A,B)$ is selected from our dataset for the connectome network ``Macaque 30".
The network has 30 nodes and has 7 input nodes.
Node 4 was selected as the input (shown as black in Fig.\,\ref{fig:signal}) and node 1 was selected as the output (shown as red in Fig.\,\ref{fig:signal}).
The distance from the input to the output node is $d=2$.
% The transfer function is
% \begin{equation}
%     G = C(sI - A)^{-1}B = \dfrac{2000 s^2 + 4.138(10^4) s + 2.14(10^5)}{s^5 + 51.1 s^4 + 1026 s^3 + 1.009(10^4) s^2 + 4.843(10^4) s + 8.933(10^4)}.
% \end{equation}
The input and the output signals are composed of the following components:
\begin{subequations}
\begin{gather}
    \text{Low Frequency input} = \sin(2\pi t), \\
    \text{High Frequency input} =0.2\sin(40\pi t), \\
    \text{Low Frequency output} = A_1 \sin(2\pi  t +\phi_1 ), \\
    \text{High Frequency output} = A_2 \sin(40\pi  t+\phi_2),
\end{gather}
\end{subequations}
where
\begin{subequations}
\begin{gather}
    A_1 = | G(2\pi \jmath) | = 1.41, \quad \phi_1 = \angle G(2\pi \jmath) = -86\degree, \\
    A_2 = | G(40\pi \jmath) | = 0.002, \quad \phi_2 = \angle G(40\pi \jmath) = -251\degree,
\end{gather}
\end{subequations}
where $\jmath=\sqrt{-1}$.
The amplitude and the phases of the output components can be seen in the Bode plot in the bottom panel of Fig.\,\ref{fig:signal} (shown as green circles for the low-frequency signal and purple squares for the high-frequency signal).
Note that magnitudes of $1.41$ and $0.002$ correspond to dB values $20\log_{10}(1.41) = 3$ and $20\log_{10}(0.002) = -54$, respectively.

\color{black}

\subsection{DC Gain} \label{sec:dcgain}

In this limit of the frequency variable approaching zero, \( \omega \to 0 \), the transfer function \( G(\jmath\omega) \) simplifies as follows:
\begin{equation*}
\begin{split}
    \text{DC Gain} & =  \lim_{\omega \to 0} 20 \log_{10} | G(\jmath\omega) | \\ 
    & =  \lim_{\omega \to 0} 20 \log_{10} | C (\jmath\omega - A)^{-1} B | \\ 
    & =  20 \log_{10} | C A^{-1} B |, 
    \end{split}
    \end{equation*}
    \begin{equation*}
    \begin{split}
    \text{Phase} = \lim_{\omega \to 0} \angle G(\jmath\omega) = \lim_{\omega \to 0} \angle C (\jmath\omega - A)^{-1} B \\ = \angle (-CA^{-1}B).
    \end{split}
\end{equation*}
In the case where an external signal is applied to node \( i \) and the response is measured at node \( j \), so that \( B = \be_i \), \( C = \be_j^\top \), the expressions simplify to
\begin{gather}
    \text{DC Gain} = 20 \log_{10} \left | [A^{-1}]_{ji} \right | = 20 \log_{10} \left | \frac{Q_{ij}}{\det(A)} \right |, \\
    \text{Phase} = \angle [A^{-1}]_{ji} =  \angle \left( \frac{Q_{ij}}{\det(A)} \right),
\end{gather}
where \( Q_{ij} \) denotes the \((j,i)\)-th cofactor of \( A \), defined as
\begin{equation*}
    Q_{ij} = (-1)^{i+j} \det(M_{ij}),
\end{equation*}
with \( M_{ij} \) the submatrix of \( A \) obtained by deleting row \( i \) and column \( j \).

\subsection{Properties of Controllability Gramian} \label{sec:propgramian}

The controllability Gramian is a symmetric and positive semi-definite matrix. 
The system is called controllable if and only if the matrix is nonsingular.
The Gramian can be directly related to the energy required to steer the system around the state space.
Assuming the initial condition for the system is at the origin and the desired final state is $\bx_f$, then the solution to the minimum energy problem is
\begin{equation}
    \text{Energy} = \dfrac{1}{2} \int_0^\infty \bu(t)^\top \bu(t) = \bx_f^\top W_c^{-1} \bx_f.
\end{equation}
Here, the entries on the main diagonal of $W_c^{-1}$ show the effort in steering the system in canonical directions $\be_i$, i.e., $\text{Energy} = \be_i^\top W_c^{-1} \be_i = [W_c^{-1}]_{ii}$.

\color{black}

\subsubsection{Trace of Continuous-Time Controllability Gramian}

We proceed under the assumption that all the eigenvalues of the matrix $A$ have negative real part. We see that an important role is played by the largest real part of the eigenvalues $a = \max_i \Re(\lambda_i (A)) < 0$.
If $a$ is large in magnitude, the trace of the continuous-time Gramian will be proportional to the number of input nodes $m$, i.e., 
\begin{equation} \label{eq:trlinear}
    \Tr (W_c) = \dfrac{m}{2 |a|},
\end{equation}
where $a = \max_i \Re(\lambda_i (A)) < 0$. 
To see this, we write $A = a I + \Delta$.
From the continuous-time Lyapunov Equation we have 
\begin{align}
\begin{split}
    A W_c + W_c A^\top + B B^\top  = & \, (a I + \Delta) W_c + W_c (a I + \Delta)^\top \\
    & \, + B B^\top \\
    = & \, 2a W_c + \Delta W_c + W_c \Delta^\top + B B^\top.
\end{split}
\end{align}
Since $\| \Delta \| \ll | a |$, then $\Delta$ acts as a small perturbation. 
Dropping the terms containing $\Delta$, we have $W_c \approx \dfrac{1}{2|a|} BB^\top$ which implies
\begin{equation} 
    \Tr (W_c) \approx \dfrac{\Tr(BB^\top)}{2 |a|}  = \dfrac{m}{2 |a|}.
\end{equation}

\begin{figure}
    \centering
    \includegraphics[width=0.9\linewidth]{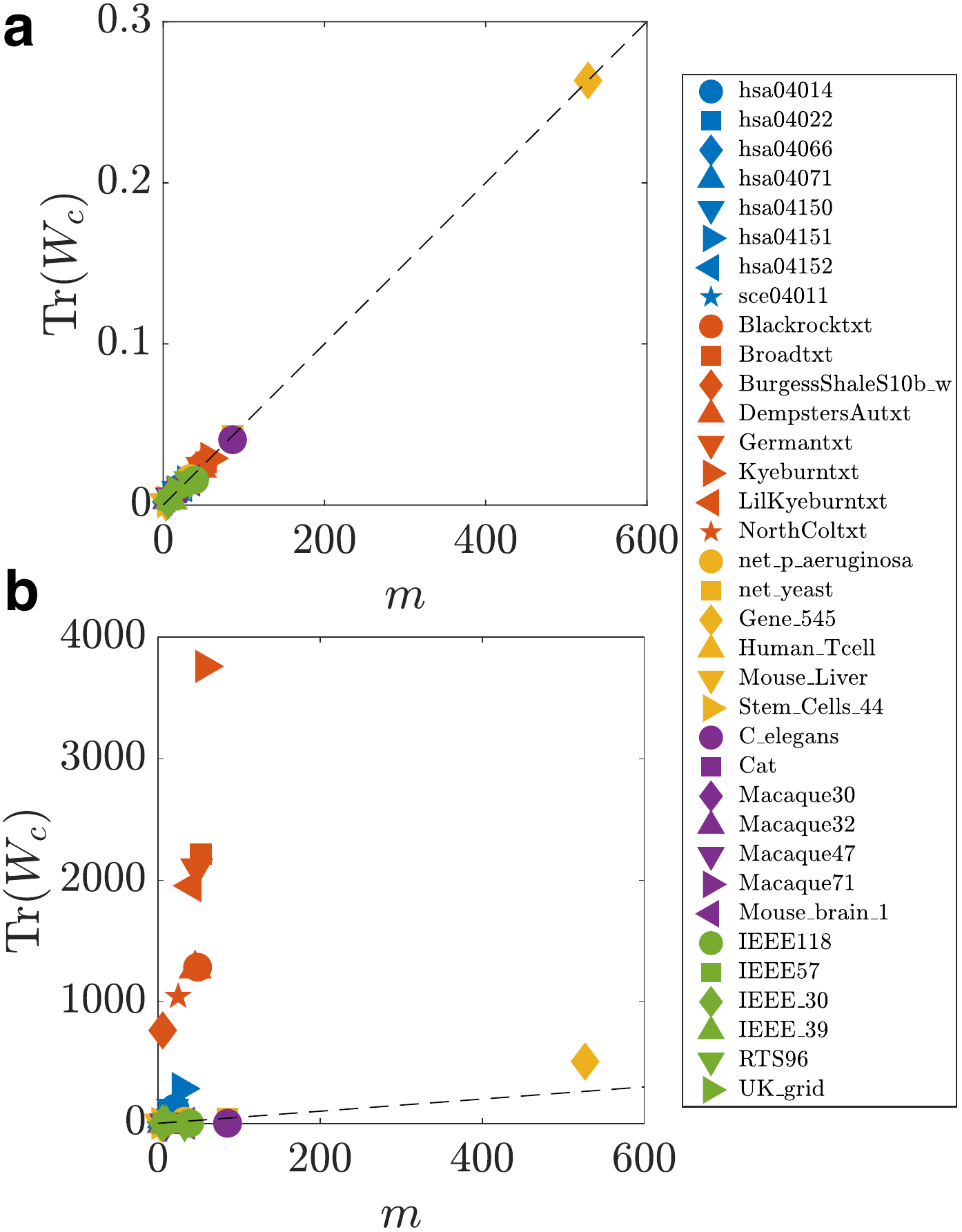} 
    \caption{\textbf{Linear relationship between the trace and the number of input nodes.}
    The trace of the infinite-horizon continuous-time Gramian $\Tr(W_c)$ for different real networks vs their number of input nodes $m$.
    In panels a and b, the matrices $A$ are shifted so that for all networks the largest real part eigenvalues is $-10^3$ and $-1$, respectively. 
    The dashed black line demonstrates the linear relationship in Eq.\,\eqref{eq:trlinear}: $\Tr (W_c) = m/2|a|  = m/2(10^{3})$. 
    The same dashed black line is maintained in Panel b for easier comparison.}
    \label{fig:trace_pinned_ct}
\end{figure}

An example of this is shown in Fig.~\ref{fig:trace_pinned_ct}, where $\Tr(W_c)$ is evaluated for selected real networks. 
We see that when $a$, the largest real part of the eigenvalues of the matrix $A$ is large in magnitude (panel a), $\Tr(W_c) \approx m/(2|a|)$, and when the magnitude is small (panel b), the trace no longer has a linear relationship with $m$.

\subsubsection{Modularity}

It was shown that properties such as the trace of the output  Gramian have the modularity property \cite{summers2015submodularity}.
Modularity is analogous to linearity, which means that each element of a subset gives an independent contribution to the function value. 
In particular case of the trace of the output Gramian, it is shown that the contribution of each column of the matrix $B$ to the trace is independent of the others, so one can check the contribution of each column individually and select the columns that provide desired contributions (e.g., the columns that minimize/maximize the trace.)

We consider the set optimization, which is the selection of a $k$-element subset of $\mathcal{V}  = \{ 1, 2, \hdots, M \}$ such that the function $f: 2^\mathcal{V} \to \mathbb{R}$ is maximized:
\begin{align} \label{eq:set}
    \max_{\mathcal{S} \subseteq \mathcal{V}, \ | \mathcal{S} | = k} \quad f(\mathcal{S}).
\end{align}
In \cite{summers2015submodularity}, it was shown that the actuator selection for the control of linear systems can be formulated as a set optimization. 

\begin{definition}[Modularity \cite{summers2015submodularity}]
    A set function $f: 2^\mathcal{V} \to \mathbb{R}$ is called modular if for all subsets $\mathcal{A}, \mathcal{B} \subseteq \mathcal{V}$, it holds that
     \begin{equation}
         f (\mathcal{A}) + f (\mathcal{B}) = f (\mathcal{A} \cup \mathcal{B}) + f (\mathcal{A} \cap \mathcal{B}).
     \end{equation}
\end{definition}
Modular set functions are similar to linear functions and have the property that each element of a subset gives an independent contribution to the function value. 
Consequently, if $f$ is modular, the optimization problem \eqref{eq:set} is easily solved by simply evaluating the set function for each element, sorting the result, and then choosing the top $k$ individual elements from the sorted list to obtain the best subset of size $k$.
It was shown that properties such as the trace of the continuous-time  Gramian have the modularity property, while the smallest eigenvalue of the  Gramian does not \cite{summers2015submodularity}.

The trace of the output  Gramian $C W_c C^\top$ also has a modularity property with respect to the rows of the output matrix $C$. 
This means that $\Tr (C W_c C^\top) = \sum_{j=1}^N C_{j,:} W_c C_{j,:}^\top$, where $C_{j,:} \in \mathbb{R}^{1 \times n}$ is row $j$ of the matrix $C$.
To see this, we use the cyclicity and linearity of the trace and write
\begin{align*}
    \Tr \left(C W_c C^\top \right) & = \Tr \left(W_c C^\top C \right) \\
    & = \Tr \left(W_c \sum_{j=1}^N C_{j,:}^\top C_{j,:} \right) \\
    & = \sum_{j=1}^N\Tr \left(W_c C_{j,:}^\top C_{j,:} \right) \\
    & = \sum_{j=1}^N\Tr \left(C_{j,:} W_c C_{j,:}^\top  \right) \\
    & = \sum_{j=1}^N C_{j,:} W_c C_{j,:}^\top .
\end{align*}

One interesting question is how to select a set of $k$ nodes in the network as input nodes so that the trace of the output  Gramian is maximized or minimized.
Through the modularity property of the trace, the trace is recorded when each node is set as the input node. Then, the $k$ nodes that have resulted in the maximum/minimum trace are selected as the optimal choices for input nodes. %{ The submodularity property implies that the `brute force' selection of input nodes provides a solution that is provably within a certain percentage of the actual best solution. However, here I think we got something stronger! Say you have assigned the output nodes and you know what is the contribution of each input-output pair then you know exactly what the best selection of input nodes will be and the value of the objective function associated; something similar can be said of the selection of output nodes, too. I propose that this is discuseed at the end of subsection D. I think we got a stronger property here than submodularity which can be exploited to solve optimization problems. }
%{Correct. The text should read the "modularity" property, not "submodularity". I fixed it.}

% {\textbf{Potential removal if current draft only discusses continuous-time:}\\
% In discrete-time setup, if the initial network is such that the spectral radius of the adjacency matrix $A$ is small, then instead of evaluating the trace for each input node, it is sufficient to calculate the network index $\alpha$ for each input node and select $k$ nodes with maximum/minimum values of $\alpha$. 
% Evaluation of $\alpha$ is much more computationally efficient than evaluating the trace, especially in large-scale networks.
% }

\subsubsection{Properties of the trace of the Output Controllability Gramian}

We define the set of input nodes $\mathcal{I}$ and the set of output nodes $\mathcal{O}$. The matrices $B$ and $C$ are uniquely defined by the set of the input nodes $\mathcal{I}$ and the set of output nodes $\mathcal{O}$, as we explain next.
The columns of the matrix $B$ are versors corresponding to the input nodes, i.e., each column has all zero entries except for a unique entry equal to one in the position corresponding to each one of the input nodes in $\mathcal{I}$.  The rows of the matrix $C$ are versors corresponding to the output nodes, i.e., each row has a unique element equal to one in the position corresponding to each one of the output nodes in the set $\mathcal{O}$.  

In accordance with the main manuscript, we define the  $\mathcal{H}_2$-norm squared of the network as an open system,
\begin{equation}
\mathcal{H}_2^2(A,\mathcal{I},\mathcal{O})=\Tr [ C W_c  C^\top ],
\end{equation}
where $W_c$ is the continuous-time controllability Gramian,
\begin{equation}
    W_c  = \int_0^{+ \infty}  e^{At} B B^\top e^{A^\top t} d t.
\end{equation}

The following two properties hold:

\textit{Additivity of the $\mathcal{H}_2$-norm with respect to the input nodes}. Consider the input node set $\mathcal{I}= \mathcal{I}_1 \cup \mathcal{I}_2 $, $\mathcal{I}_1 \cap \mathcal{I}_2 = \emptyset $, then
\begin{equation}
\mathcal{H}_2^2(A,\mathcal{I},\mathcal{O})=\mathcal{H}_2^2(A,\mathcal{I}_1,\mathcal{O})+\mathcal{H}_2^2(A,\mathcal{I}_2,\mathcal{O}). \label{add_inp}
\end{equation}

The proof can be found in \cite{summers2015submodularity}. 
From \eqref{add_inp} it follows trivially that
\begin{equation}
\mathcal{H}_2^2(A,\mathcal{I},\mathcal{O})=\sum_{i \in \mathcal{I}} \mathcal{H}_2^2(A,i,\mathcal{O}).
\end{equation}

\textit{Additivity of the $\mathcal{H}_2$-norm with respect to the output nodes}. Consider the output node set $\mathcal{O}= \mathcal{O}_1 \cup \mathcal{O}_2 $, $\mathcal{O}_1 \cap \mathcal{O}_2 = \emptyset $, then
\begin{equation} \label{add_out}
\mathcal{H}_2^2(A,\mathcal{I},\mathcal{O})=\mathcal{H}_2^2(A,\mathcal{I},\mathcal{O}_1)+\mathcal{H}_2^2(A,\mathcal{I},\mathcal{O}_2).
\end{equation}

The proof follows from these two facts: (i) the output  Gramian with output set $\mathcal{O}_i$ is equal to a minor of the  Gramian obtained by selecting the rows/columns of the  Gramian corresponding to the nodes in $\mathcal{O}_i$ \cite{klickstein2017energy} and (ii) the trace of a square matrix is equal to the sum of the elements on the main diagonal.
From \eqref{add_inp} and \eqref{add_out} it follows trivially that
\begin{equation}
\mathcal{H}_2^2(A,\mathcal{I},\mathcal{O})=\sum_{o \in \mathcal{O}} \mathcal{H}_2^2(A,\mathcal{I},o)= \sum_{i \in \mathcal{I}} \sum_{o \in \mathcal{O}} \mathcal{H}_2^2(A,i,o),
\end{equation}
where $\mathcal{H}_2^2(A,i,o)$ is the individual contribution of input node $i$ and output node $o$ to the overall  $\mathcal{H}_2$-norm of the network.

\color{black}

\subsection{Dynamical models for each network type} \label{methods:models}

The matrix $A$ considered in this work is the Jacobian of the network dynamics close to a stable fixed point, for each type of network. Therefore, it depends on both the specific structure of the network and the dynamics that it supports. Here, we describe the dynamical models we used for each type of network dynamics considered in this study. 
\subsubsection{Food webs}
We employ the generalized Lotka-Volterra model, which is a well-known model of competition and mutualism among interacting species \cite{asllani2018structure,may1972will,allesina2012stability,coyte2015ecology}.
The governing equations of this model are:
\begin{equation}
    \dfrac{d z_i (t)}{dt} = z_i (t) \left( r_i -s_i z_i (t) \textcolor{black}{+} \sum_{j\neq i} \tilde{A}_{ij} z_j (t) \right),
\end{equation}
$i = 1, \hdots, N$, where $z_i (t)$ is the population of species $i$, $r_i$ are the intrinsic rates of birth if $r_i > 0$, or death if $r_i < 0$. 
The positive constants $s_i$ represent the finite carrying capacity of the ecosystem (limited resources) and prevent species $i$ from growing indefinitely.
The interaction among species is given by the adjacency matrix $\tilde{A}=[\tilde{A}_{ij}]$, which has zeros on its main diagonal, i.e., $\tilde{A}_{ii}=0, \forall i$.

Following \cite{asllani2018structure}, we assume $s_i = 1, \forall i$, for simplicity.
For a proper choice of $r_i$, the point $z_i^*=1, \forall i$, is the fixed-point of the system (if $r_i = \sum_{j\neq i}\tilde{A}_{ij}, \forall i$).
Under these conditions, the Jacobian matrix $A = [A_{ij}]$ evaluated at the fixed-point $x_i^*=1$ becomes
\begin{equation}
    A_{ij}=\begin{cases}
         -\sum_{k\neq i} \tilde{A}_{ik}, & \quad \text{if} \ \ i=j, \\
        \\
        -\tilde{A}_{ij}, & \quad \text{if} \ \ i\neq j.
    \end{cases}
\end{equation}
The Jacobian $A$ has all non-positive real-part eigenvalues for networks with non-negative weights (this follows from the Gershgorin circle theorem).
In practice, it may sometimes happen that the matrix $A$ has zero eigenvalues, which is undesired when calculating the Gramian (the Gramian will not converge).
Therefore, we apply the spectral shift  $A \leftarrow A-cI$   and properly choose $c$ so that the largest real part eigenvalue of the matrix $A$ is $-1$. The input nodes are identified and the matrix 
$B$ is constructed as detailed in Supplementary Note 4A. We then perform our analysis for the pair $(A,B)$, where the matrix $B$ is constructed using knowledge of the particular selection of the input nodes from the empirical datasets.

\subsubsection{Connectomes}\label{sec:connectomes}

Following \cite{PetkoskiJirsa2019,koller2024human}, we use the Kuramoto model~\cite{kuramoto1984chemical}
to describe the interaction dynamics within connectome networks,  
\begin{align}
    \dfrac{d \theta_i (t)}{dt}  = \omega_i - \frac{K}{N}\sum_{j=1}^N \tilde{A}_{ij} \sin(\theta_i(t)-\theta_j(t)), 
\end{align}
$i = 1, \hdots, N$, where $\theta_i\in(-\pi,\pi]$ and $\omega_i\in\mathbb{R}$ correspond, respectively, to the phase and the natural frequency at node $i$\,. 
The $N$-dimensional matrix $\tilde{A}=[\tilde{A}_{ij}]$ is the adjacency matrix, describing the network topology, and $N$ is the number of nodes.
When $\kappa=\frac{K}{N}$ is large enough, the system typically reaches a phase-locking state where all the phases evolve at the same angular frequency given by $\Omega = N^{-1}\sum_{j=1}^N \omega_j$\,.
The phase-locked state is defined as when $d\theta_i (t)/dt = \Omega, \forall i$. Therefore, the phased-locked state for oscillator $i$ is $\theta_i^*(t)=\theta_i^*(0)+\Omega t$.

The Jacobian of the system $A = [A_{ij}]$ linearized about the phase-locked state is,
\begin{equation}
    A_{ij} = \begin{cases}
         -\sum_{k} \tilde{A}_{ik}\cos(\theta_i^*(0) -\theta_k^*(0)), & \quad \text{if} \ \ i=j, \\
        \\
        \tilde{A}_{ij}\cos(\theta_i^*(0) -\theta_j^*(0)), & \quad \text{if} \ \ i\neq j.
    \end{cases}
\end{equation}
In this case, the Jacobian $A$ has some of the entries in its rows equal zero, thus it has at least one eigenvalue equal to 0. 
Since our methodology requires asymptotic stability,  we apply the spectral shift $A \leftarrow A-cI$ and properly choose $c$ so that the largest real part eigenvalue of the matrix $A$ is $-1$. The input nodes are identified and the matrix 
$B$ is constructed as detailed in Supplementary Note 4B. We then perform our analysis for the pair $(A,B)$, where the matrix $B$ is
constructed from knowledge of the particular selection
of the input nodes derived from the empirical datasets.

In our simulations, we randomly select the natural frequencies $\omega_i$ from the standard uniform distribution (between 0 and 1).
We set $K=N$, so that $\kappa = 1$.

\subsubsection{Power grids}
We consider a simplified swing dynamics for the voltage angles, including losses on the transmission lines. The dynamics obeys the equations~\cite{machowski2020power},
\begin{align}
    \dfrac{d \theta_i (t)}{dt} &= P_i - \sum_{j\neq i} B_{ij}\sin(\theta_i-\theta_j) + G_{ij}[1-\cos(\theta_i-\theta_j)],
\end{align}
$i = 1, \hdots, N$, where $\theta_i\in(-\pi,\pi]$ is the voltage phase and $P_i$ is the injected/consumed power at bus (node) $i$\,. The transmission line between buses $i$ and $j$ have susceptance $B_{ij}$, and conductance $G_{ij}$\,. The Jacobian matrix $A$ is then obtained by linearizing the dynamics around a stable operational state of the grid which corresponds to a phase-locked state similar to Sec.~\ref{sec:connectomes}. Each fixed point depends on the topology of the grid and the dispatch of injected/consumed power $P_i$\,. The entries of the Jacobian are,
\begin{equation}\label{eq:swing}
    A_{ij}=\begin{cases}
         -\sum_{k\neq i} \tilde{A}_{ik}\cos(\theta_i^*-\theta_k^* + \gamma_{ij}), & \quad \text{if} \ \ i=j, \\
        \\
        -\tilde{A}_{ik}\cos(\theta_i^*-\theta_j^* + \gamma_{ij}), & \quad \text{if} \ \ i\neq j,
    \end{cases}
\end{equation}
where $\tilde{A}_{ij} = \sqrt{B_{ij}^2 + G_{ij}^2}$\,, and $\gamma_{ij} = \arctan(-G_{ij}/B_{ij})$\,. 
As the rows of the matrix $A$ in (\ref{eq:swing}) sum to zero, it has at least one vanishing eigenvalue. Since our methodology requires asymptotic stability,  we apply the spectral shift $A \leftarrow A-cI$ and properly choose $c$ so that the largest real part eigenvalue of the matrix $A$ is $-1$. The input nodes are identified and the matrix 
$B$ is constructed as detailed in Supplementary Note 4C. We then perform our analysis for the pair $(A,B)$, where the matrix $B$ is
constructed from knowledge of the particular selection
of the input nodes derived from the empirical datasets.

\subsubsection{Genetic and Pathway networks}

We use the regulatory model (Michaelis-Menten model \cite{karlebach2008modelling}) to capture pathways and genetic dynamics. 
The model obeys the equations,
\begin{equation}
    \dfrac{d z_i (t)}{dt} = -f_i z_i^a (t) + \kappa \sum_{j=1}^N \tilde{A}_{ij}  \dfrac{z_j^h (t)}{1+z_j^h (t)}, 
\end{equation}
$i = 1,\hdots, N$, where $z_i (t)$ is the gene expression at time $t$, the local dynamics $-f_i z_i^a (t)$ captures biochemical processes (where $a$ depends on the process), the matrix $\tilde{A}=[\tilde{A}_{ij}]$ is the weighted adjacency matrix of the network, and the term $z_j^h (t) / (1+z_j^h (t))$ describe genetic activation, where $h$ modulates the saturation of the term and is associated with the level of cooperation in gene regulation.
For further details of the model and its nonlinear analysis, see \cite[Supplementary Section 4.2]{meena2023emergent}.

The Jacobian of the system linearized about the active fixed-point (a fixed-point other than the origin) $\bz^*=[z^*_i]$ is $A = [A_{ij}]$ where
\begin{equation}
    A_{ij} = \begin{cases}
         -f_ia {z_i^*}^{a-1}, & \quad \text{if} \ \ i=j, \\
        \\
        \kappa \tilde{A}_{ij} \dfrac{h{z^*}_j^{h-1}}{(1+{z^*}_j^{h})^2}, & \quad \text{if} \ \ i\neq j.
    \end{cases}
\end{equation}
As with other models, we apply the spectral shift $A \leftarrow A-cI$ and properly choose $c$ so that the largest real part eigenvalue of the matrix $A$ is $-1$. We then perform our analysis for the pair $(A,B)$, where the matrix $B$ is constructed from knowledge of the particular selection of the input
nodes derived from the empirical datasets, see Supplementary Notes 4D and 4E.

In our simulations, we set $a=2$, $h=1$, $\kappa = 1$, and sample $f_i$ from the standard uniform distribution (between 0 and 1).
With this choice of parameters, an active fixed-point (a fixed-point other than the origin) exists and is stable.

\color{black}

\section*{Data availability}
All data generated or analyzed during this study are included in this published article (and its supplementary information files).

\section*{Code availability}
The source code for the numerical simulations presented in the paper will be made available upon request, as the code is not required to support the main results reported in the manuscript.

\section*{Acknowledgements}
We acknowledge support from grants AFOSR FA9550-24-1-0214 and Oak Ridge National Laboratory 006321-00001A. Wai Lim Ku was supported by the National Institutes of Health grant K22HL153477

\section*{Author Contributions}
Amirhossein Nazerian worked on the theory and numerical simulations. {Malbor Asllani and Melvyn Tyloo worked on the theory and the data analysis.} Wai Lim Ku provided assistance with biological data. Francesco Sorrentino worked on the theory and supervised the research. 
All authors contributed to writing the paper.

\section*{Competing Interests Statement}
The authors declare no competing interests.

\end{document}